\documentclass[a4paper,11pt]{article}
\pdfoutput=1 % if your are submitting a pdflatex (i.e. if you have
\usepackage{jheppub}
\usepackage{color} 
\usepackage{float}
\usepackage{graphicx}
\usepackage{amsmath} 
\usepackage{ulem}

\usepackage{amsmath,graphicx,amssymb,subfigure,bbm}

\usepackage[utf8]{inputenc}
\usepackage[T1]{fontenc}

\usepackage{bbold}

\usepackage[all]{xy}

\usepackage{verbatim}

\usepackage{xcolor}

\colorlet{darkred}{red!80!black}

\newcommand{\be}{\begin{equation}}
\newcommand{\ee}{\end{equation}}
\newcommand{\bea}{\begin{eqnarray}}
\newcommand{\eea}{\end{eqnarray}}

\usepackage{soul}

\def\beq{\begin{equation}}
\def\eeq{\end{equation}}
\def\beqa{\begin{eqnarray}}
\def\eeqa{\end{eqnarray}}

\begin{document}

\title{A dynamical holographic QCD model for spontaneous chiral symmetry breaking and confinement}

%\title{\boldmath Holographic description of spontaneous chiral symmetry breaking in a confined background}

\author[a]{Alfonso Ballon-Bayona,}
\author[b]{Tobias Frederico,}
\author[c, d]{Luis A. H. Mamani,}
\author[b]{and Wayne de Paula}
\affiliation[a]{Instituto de F\'{i}sica, Universidade
Federal do Rio de Janeiro, \\
Caixa Postal 68528, RJ 21941-972, Brazil.}       
\affiliation[b]{Instituto Tecnol\'ogico de Aeron\'autica, DCTA, 12228-900 S\~ao Jos\'e dos Campos, Brazil}
\affiliation[c]{Centro de Ci\^encias Exatas e Tecnol\'ogicas,
Universidade Federal do Rec\^oncavo da Bahia,\\
Rua Rui Barbosa, 710, 44380-000, Cruz das Almas, Bahia, Brazil}
\affiliation[d]{Laborat\'orio de Astrof\'{\i}sica Te\'orica e Observacional,\\ 
Departamento de Ci\^encias Exatas e Tecnol\'ogicas,
Universidade Estadual de Santa Cruz,\\ 45650-000, Ilh\'eus, Bahia, Brazil}
\emailAdd{aballonb@if.ufrj.br}
\emailAdd{tobias@ita.br}
\emailAdd{luis.mamani@ufrb.edu.br}
\emailAdd{wayne@ita.br}

\abstract{
In this paper, we present a holographic realization of spontaneous chiral symmetry breaking and confinement. The latter is realized by building a solution of 5d Einstein-dilaton gravity leading to a confining quark antiquark potential. The 4d currents and the quark mass operator associated with chiral symmetry breaking and  creation of meson states are mapped to 5d fields whose dynamics is given by a non-Abelian Higgs action. We introduce a non-minimal dilaton coupling to the tachyon potential which has two parameters, one of them controlling the presence or absence of spontaneous chiral symmetry breaking and the other controlling the sign of the chiral condensate as the quark mass grows. We calculate the masses of vector, scalar, axial-vector and pseudoscalar mesons, focusing on the effect of chiral symmetry breaking on the spectrum. In the chiral limit we identify the emergence of a massless state in the pseudoscalar meson spectrum, i.e., the pion.  We calculate all meson decay constants and confirm that the pion satisfies the Gell-Mann-Oakes relation in the light quark regime.
}

\maketitle
\flushbottom

\section{Introduction}

The investigation of internal symmetries in physical systems is an active research area. In particle physics, it was initiated with the discovery of the  neutron and proton, where the concept of isospin SU$(2)$ symmetry 
was motivated by the fact of they have aproximately the same mass and interaction \cite{RevModPhys.81.1015} . The concept of spontaneous symmetry breaking (SSB) applied in particle physics, was first introduced in \cite{PhysRev.127.965,PhysRev.128.2462}. As a consequence of SSB, massless particles, called Nambu-Goldstone bosons, arise in the theory \cite{PhysRev.127.965}. In quantum chromodynamics (QCD), the pions are pseudo-Nambu-Goldstone bosons, arising when the chiral-symmetry group, SU(3)$_L\times$SU(3)$_R$, is spontaneously broken in the low-energy regime, where perturbation techniques are not reliable.  Non-perturbative techniques are required in order to extract relevant information of QCD in the low-energy regime. Meanwhile, from the field theory point of view, is it believed that a confining theory shall induce the formation of a non-trivial quark condensate, see for example \cite{Tong2018}. In the large N limit of QCD, a connection between confinement and spontaneous chiral symmetry was obtained  by Coleman and Witten \cite{Coleman:1980mx}.

Using the AdS/CFT dictionary  \cite{Maldacena:1997re,Maldacena:2000yy, Gubser:1998bc, Witten:1998qj}, several groups around the world have investigated 5d gravitational theories dual to 4d gauge theories in the large N limit that are UV conformal in order to describe the phenomenology of hadrons, confinement and chiral symmetry breaking  \cite{BoschiFilho:2002ta, BoschiFilho:2002vd, Erlich:2005qh, Karch:2006pv, Karch:2002sh, Evans:2004ia, Sakai:2004cn, deTeramond:2005su,Sakai:2005yt, Andreev:2006ct,Grigoryan:2007vg,Gursoy:2007er, Cherman:2008eh,  dePaula:2008fp, Colangelo:2008us, Vega:2008af, Abidin:2009hr,Abidin:2009aj, dePaula:2009za, Bianchi:2010cy, Gherghetta:2009ac, Jarvinen:2011qe, Li:2013oda, Arean:2013tja,Alho:2013dka,Ballon-Bayona:2014oma,Chelabi:2015gpc,Ballon-Bayona:2017sxa, Ballon-Bayona:2017bwk, Ballon-Bayona:2020qpq, Ballon-Bayona:2021ibm}, see also references therein. At finite temperature, density and magnetic field it was also investigated the equation of state, phase diagram and transport coefficients \cite{Gubser:2008yx,Gursoy:2008za,Bigazzi:2009bk,Bigazzi:2011it,He:2013qq,Ballon-Bayona:2013cta,Alho:2013hsa,Bigazzi:2014qsa,Dudal:2014jfa,Rougemont:2015oea,Knaute:2017opk,Critelli:2017oub,Ballon-Bayona:2017dvv,Gursoy:2017wzz,Braga:2018zlu,Arefeva:2020byn, Rannu:2022fxw, Arefeva:2022avn, Grefa:2022fpu, Grefa:2021qvt,dePaula:2020bte,Li:2023mpv}, see also references therein. Many results obtained using the  holographic techniques are in either quantitative or qualitative agreement with either experimental data or lattice QCD.  

In this paper, we present a holographic dynamical model that is confining, by the Wilson loop criterion, and leads to spontaneous chiral symmetry breaking. Within the  bottom-up approach, this subject was investigated in the holographic hard wall model \cite{Erlich:2005qh} and similar setups \cite{Alho:2013dka}. Meanwhile, in the soft wall model \cite{Karch:2006pv}, it was suggested that a non-linear term in the potential was needed to describe spontaneous chiral symmetry breaking. The first attempt to describe spontaneous chiral symmetry breaking in the soft wall model was done in 
 \cite{Gherghetta:2009ac}. However, the quark mass and chiral condensate in that model were independent parameters and it was not possible to see the evolution of the chiral condensate as a function of the quark mass. In turn, in \cite{Ballon-Bayona:2021ibm} it was observed the emergence of the pion and the evolution of the chiral condensate as a function of the quark mass, which we call, dynamical quarks. However, those results were obtained neglecting the confining properties in the gluon sector, which can be done considering Einstein-dilaton equations in the gravitational side of the duality, see for example \cite{Ballon-Bayona:2017sxa}. Here, we do a step forward, describing spontaneous chiral symmetry breaking and confinement in the same setup. Basically, we extend the analysis of \cite{Ballon-Bayona:2021ibm}, this time considering a gravitational background obtained by solving the Einstein-dilaton equations. We build an analytical solution for the metric and the dilaton field, inspired by the proposal of \cite{dePaula:2008fp}, and show that the background leads to a confining quark antiquark potential. We follow the procedure usually referred in the literature as the potential reconstruction method, where one postulates the form of the warp factor and/or the dilaton field and then one uses the remaining Einstein equation to reconstructs the dilaton potential. The main difference in relation to previous results in the literature is that we codify all the relevant ingredients to describe spontaneous chiral symmetry breaking on the dilaton coupling to the tachyon potential.

The paper is organized as follows: In subsection \ref{Sec:Background} we investigate the gluon sector of the holographic model based on five dimensional Einstein-dilaton gravity. We present a novel analtyical solution for the warp factor and dilaton field. We also investigate the criterion for confinement, namely we show that  the relevant function associated with the warp factor in the string frame displays a global minimum different from zero.  In subsection \ref{Sec:XSB} we present the five-dimensional non-Abelian Higgs action describing the chiral condensate and the mesons in the dual field theory and we postulate the form of the dilaton coupling controlling spontaneous chiral symmetry breaking. In section \ref{Sec:Mesons} we make a brief summary of the relevant fields and equations describing the spectrum of the vector, scalar, axial-vector and pseudoscalar mesons. We solve the eigenvalue problem for each sector and compare our results for the meson masses against the available experimental data. In section \ref{Sec:Decay} we calculate the meson decay constants and analyze their dependence on the quark mass. In the end of that section we investigate the Gell-Mann-Oakes-Renner relation (GOR) numerically. Our main conclusions are displayed in Section \ref{Sec:Conclusions}. Additional material is presented in the appendices.

\section{Holographic model}

\subsection{The Einstein-Dilaton background and confinement}
\label{Sec:Background}

The starting point for descrbing the gluon sector of  holographic QCD models is the five-dimensional Einstein-dilaton action
\noindent
\begin{equation}\label{EqAction}
S= \sigma \int d^{5}x\,\sqrt{-g}\Big [
R -\frac43\,\partial^{m}\Phi\partial_{m}\Phi + \ell^{-2} V(\Phi) \Big ]\ ,
\end{equation}
\noindent
where $\Phi$ is the dilaton field,  $V(\Phi)$ its potential, and 
$\sigma = M_p^3N_c^2$ is the effective 5d gravitational coupling \footnote{$M_p$ is the effective Planck constant and $N_c$ is the number of colors in the four-dimensional gauge theory.}. In this work we will fix the  AdS radius as $\ell=1$. From the variational principle, one get two set of differential equations from \eqref{EqAction}; the Einstein equations and the (generalized) Klein-Gordon equation: 
\noindent 
\begin{align}
&G_{mn} =\frac{1}{2 \sigma}T_{mn},  \label{EqEinstein} \\
\frac43\,\frac{1}{\sqrt{-g}}
&\partial_{m}\left(\sqrt{-g} \, g^{mn}\partial_n\Phi\right)
+\frac12\,\frac{dV}{d\Phi} =0, \label{Eq:KG}
\end{align}
\noindent
where $G_{mn}$ is the Einstein tensor and $T_{mn}$ the 
energy-momentum tensor defined by 
\noindent
\begin{equation}
T_{mn}= - \frac{2}{\sqrt{-g}} \frac{\delta S_M}{\delta g^{mn}} = 
\sigma \left[\frac83\,\partial_{m}\Phi\partial_n\Phi
+g_{mn} {\cal L}_{\Phi} \right],
\end{equation}
\noindent
and $S_{M} = \sigma \int d^5 x \sqrt{-g} {\cal L}_{\Phi}$ is the matter action due to the scalar field. In holographic QCD models one considers the following ansatz for the five-dimensional metric and dilaton fields
\noindent
\begin{align}\label{Eq:metric}
ds^2 &=e^{2A(u)} \left(du^2 + \eta_{\mu\nu} \, d \tilde x^\mu d \tilde x^\nu \right) \, , \nonumber \\
\Phi &= \Phi(u) \, ,
\end{align}
\noindent
where we introduced the dimensionless coordinates $u$ and $\tilde x$ related to the dimensionful coordinates $z$ and $x$ by  $u=\Lambda z$ and $\tilde x = \Lambda x$. The parameter $\Lambda$ will be interpreted as in infrared mass gap in the dual theory \footnote{In fact, a non-vanishing $\Lambda$ is responsible for conformal symmetry breaking. }.   The tensor components $\eta_{\mu\nu}$ correspond to the Minkowski metric, $A(u)$ is the warp factor in Einstein frame and Greek letters take the values $\mu,\nu=0,1,2,3$. Plugging \eqref{Eq:metric} into \eqref{EqEinstein} one finds the following independent field equations
\noindent
\begin{subequations}
\begin{align}
A'^2-A''-\frac{4}{9}\Phi'^{\,2}=&\,0,\label{Eq:EinsteinDil}\\
 (\partial_u + 3 A') A' -\frac{e^{2A}}{3}V=&\,0 ,\label{Eq:Potential}
\end{align}
\end{subequations}
\noindent
where $'$ means $d/du$. It is worth pointing out that one may write  the warp factor equation \eqref{Eq:EinsteinDil} in a linear form defining the new function $\zeta_1=e^{-A}$ \cite{Ballon-Bayona:2017sxa},
\noindent
\begin{equation}
\zeta_1''-\frac{4}{9}\Phi'^{\,2}\,\zeta_1=0.
\end{equation}
\noindent
This equation seems the harmonic oscillator differential equation with frequency given by $\omega^2\propto -\Phi'^2$. The advantage of this function is that in the conformal limit, i.e., in the limit of vanishing dilaton field, it reduces to $\zeta_1=u$. In this work, however, we shall work with the original warp factor $A(u)$. 

The system of coupled differential Eqs.~\eqref{Eq:EinsteinDil} and \eqref{Eq:Potential} may be solved following three approaches. In the first approach one postulates the dilaton field using phenomenological constraints and find the warp factor solving Eq.~\eqref{Eq:EinsteinDil} and reconstructs the dilaton potential solving Eq.~\eqref{Eq:Potential}. This approach was followed in several works in the literature, see for instance \cite{ dePaula:2009za, dePaula:2008fp,Li:2013oda, Li:2014dsa, Li:2014hja,Ballon-Bayona:2017sxa, Ballon-Bayona:2018ddm, Ballon-Bayona:2021tzw, Mamani:2019mgu} and references therein. In the second approach one postulates the warp factor considering phenomenological constraints, then calculates the dilaton field using Eq.~\eqref{Eq:EinsteinDil} and reconstructs the dilaton potential using Eq.~\eqref{Eq:Potential}. This approach was followed by several works in the literature, see for instance \cite{Cai:2012xh, He:2013qq, Cremonini:2012ny, Mamani:2020pks, Ballon-Bayona:2020xls} and references therein. The first two approaches where the dilaton potential is reconstructed from the dilaton and warp factor are usually referred as the potential reconstruction method \cite{Cai:2012xh}. In the third approach one builds the dilaton potential and solve the coupled equations \eqref{Eq:EinsteinDil} and \eqref{Eq:Potential}, see for instance \cite{Gubser:2008yx, Gursoy:2007er, Gursoy:2007cb} and references therein.

\begin{figure}[ht!]
\centering
\includegraphics[width=7cm]{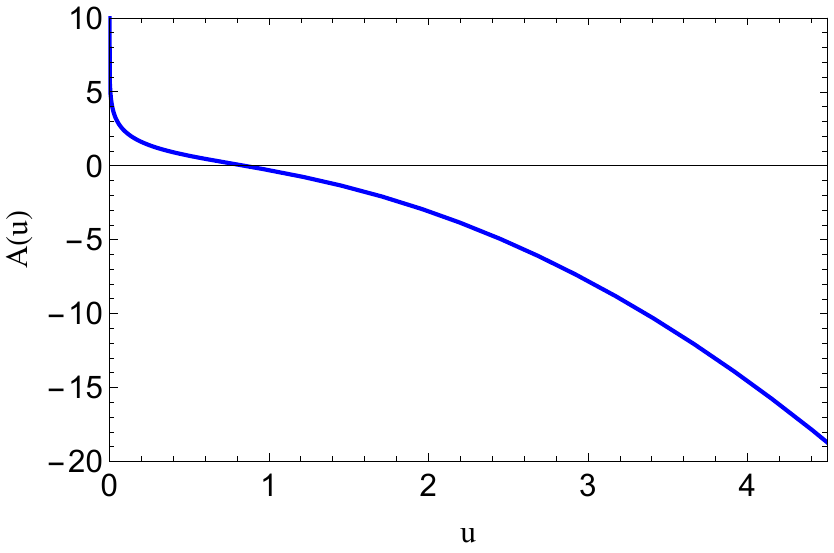}\hfill
\includegraphics[width=7cm]{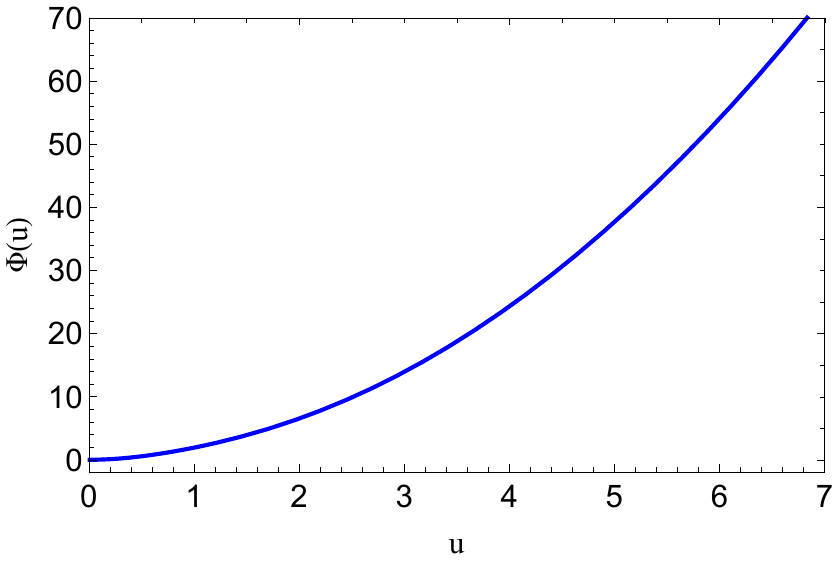}
\caption{
Left panel: The warp factor $A(u)$ in Einstein frame. Right panel: The dilaton field $\Phi(u)$.
}
\label{Fig:WarpDilaton2}
\end{figure}

In this work we will consider a model where both the warp factor and dilaton fields are analytical functions. As we want to describe confinement properties for the gluon sector of the 4d field theory, the 5d warp factor must have an specific asymptotic behavior far from the boundary, i.e. in the infrared regime (IR) \cite{Gursoy:2007er}, see also \cite{Li:2013oda, Ballon-Bayona:2017sxa}. In turn, the asymptotic behavior of the dilaton in the IR is determined by the asymptotic behaviour of the warp factor in the IR, because they are coupled through the Einstein equation \eqref{Eq:EinsteinDil}. On the other hand, close to the boundary, the  5d metric should become AdS in order to apply the holographic dictionary. Putting together these requirements on the warp factor, it must have the following asymptotic form:
\noindent
\begin{align}
    A(u)=\,&-u^2+\ln{u}+\cdots,\qquad u\to \infty \, ,\\
    A(u)=\,&-\ln{u}+\cdots,\qquad u\to 0 \, .
\end{align}
\noindent
One can build an interpolation function containing the desired asymptotic expansions. Thus, we consider the simplest interpolation function for the warp factor,
\noindent
\begin{equation}\label{Eq:WarpFactorEinstein}
\begin{split}
A(u)=\,&-\ln{u}-u^2+\ln{(1+u^2)}.
\end{split}
\end{equation}
\noindent
The asymptotic behavior of the warp factor, close to the boundary, is given by
\noindent
\begin{equation}\label{Eq:WarpEinsUV}
\begin{split}
A(u)=\,&-\ln{u}-\frac{1}{2}u^4+\frac{1}{3}u^6+\cdots,\qquad u\to 0 \, ,
\end{split}
\end{equation}
\noindent
while the asymptotic behavior in the IR is
\noindent
\begin{equation}\label{Eq:WarpEinsIR}
\begin{split}
A(u)=\,&-u^2+\ln{u}+\frac{1}{u^2}-\frac{1}{2\,u^4}+\frac{1}{3u^6}-\frac{1}{4u^8}+\cdots,\qquad u\to \infty \, .
\end{split}
\end{equation}
\noindent
A plot of the warp factor is displayed in the left panel of Fig.~\ref{Fig:WarpDilaton2}. In turn, the dilaton field is obtained solving one of the Einstein's equations, i.e., Eq.~\eqref{Eq:EinsteinDil}, which can be written in the form
\noindent
\begin{equation}
\frac{4}{9}\left(\Phi'\right)^2=\left(A'\right)^2-A''.
\label{dilatonwarp}
\end{equation}
\noindent
For this model, we can solve Eq. (\ref{dilatonwarp}) in an exact form. Then, we fix the integration constant using the asymptotic solution close to the boundary. Thus, the dilaton field is given by
\noindent
\begin{align} \label{Eq:DilatonSol}
\Phi(u) &= -\frac{3}{8 \sqrt{2}} \Big \{ \sqrt{2} \sinh ^{-1}\left(\frac{4 u^2+3}{\sqrt{31}}\right)-4 \sqrt{2 u^4+3 u^2+5} \nonumber \\
&+8 \tanh ^{-1}\left(\frac{7-u^2}{4 \sqrt{2 u^4+3 u^2+5}}\right) \Big \} - \Phi_0 \, ,
\end{align}
where we have defined the constant
\begin{equation}
\Phi_0 = -\frac{3 \left(-4 \sqrt{5}+\sqrt{2} \sinh ^{-1}\left(\frac{3}{\sqrt{31}}\right)+8 \tanh ^{-1}\left(\frac{7}{4 \sqrt{5}}\right)\right)}{8 \sqrt{2}} 
\approx -0.0535173 \, ,
\end{equation}
\noindent
that allows us to set the dilaton field to zero at $u=0$. By expanding the dilaton close to the boundary we get
\noindent
\begin{equation}
\begin{split}
\Phi(u)=\,&\frac{3}{2}\sqrt{\frac{5}{2}}\,u^2-\frac{21}{8\sqrt{10}}u^4+\cdots, \qquad u\to 0 \, .
\end{split}
\end{equation}
\noindent
In the same way, we get the asymptotic dilaton field in the IR 
\noindent
\begin{equation}\label{Eq:PHI_IR}
\begin{split}
\Phi(u)=\,&\Phi^{\infty}+\frac{3}{2}u^2-\frac{3}{4}\ln{u}-\frac{117}{64u^2}+\cdots,\qquad u\to \infty \, ,
\end{split}
\end{equation}
\noindent
where
\begin{align}
\Phi^{\infty}  &=  - \frac{3}{16} \log \left(\frac{64}{31} \right)-\Phi_0 +\frac{9}{8}+\frac{3 \coth ^{-1}\left(4 \sqrt{2}\right)}{\sqrt{2}} \, . 
\end{align}    
The dilaton profile $\Phi(u)$ is displayed on the right panel of Fig.~\ref{Fig:WarpDilaton2}. As can be seen from the figures, the warp factor and dilaton fields are  smooth functions of the holographic coordinate $u$.

\begin{figure}[ht!]
\centering
\includegraphics[width=7cm,height=4.5cm]{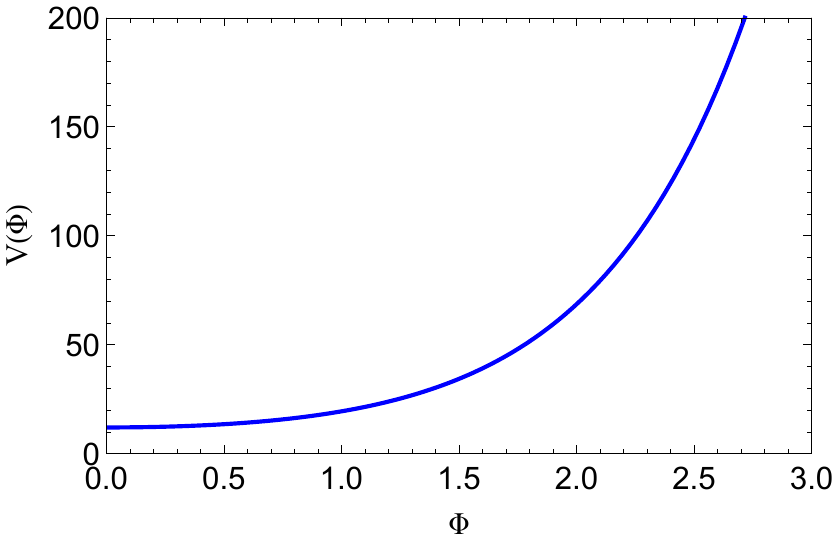}
\caption{
The dilaton potential as a function of the dilaton field.
}
\label{Fig:PhiV}
\end{figure}

In addition, we reconstruct the dilaton potential $V(\Phi)$ using equation \eqref{Eq:Potential} and present it in Fig.~\ref{Fig:PhiV}. In the limit $\Phi \to 0$ one gets $V=12$, which is related to the 5d negative cosmological constant associated with asymptotically AdS spacetimes \footnote{Namely, one obtains $R  + 12/\ell^2$ and recovers 5d Einstein gravity with a negative cosmological constant.}.

Let us turn our attention to the warp factor in the string frame, which can be calculated using the relation
\noindent
\begin{equation} \label{Eq:stringwarpfactor}
A_s=A+\frac{2}{3}\Phi. 
\end{equation}
\noindent
Using our results for $A(u)$ and $\Phi(u)$
we find that close to the boundary $A_s(u)$ behaves as
\noindent
\begin{equation}\label{Eq:WarpStringUV}
\begin{split}
A_{s}(u)=\,&-\ln{u}+\sqrt{\frac52}\, u^2-\left(\frac12+\frac{7}{4\sqrt{10}}\right)u^4+\cdots,\qquad u\to 0
\end{split}
\end{equation}
\noindent
while far from the boundary (IR region)  $A_s(u)$ behaves as
\noindent
\begin{equation}\label{Eq:WarpStringIR}
\begin{split}
A_{s}(u)=\,&\frac23\Phi^{\infty}+ \frac12 \ln u -\frac{7}{32u^2}+\frac{121}{256u^4}+\cdots.\qquad u\to \infty
\end{split}
\end{equation}
\begin{figure}[hbt!]
\centering
\includegraphics[width=7cm]{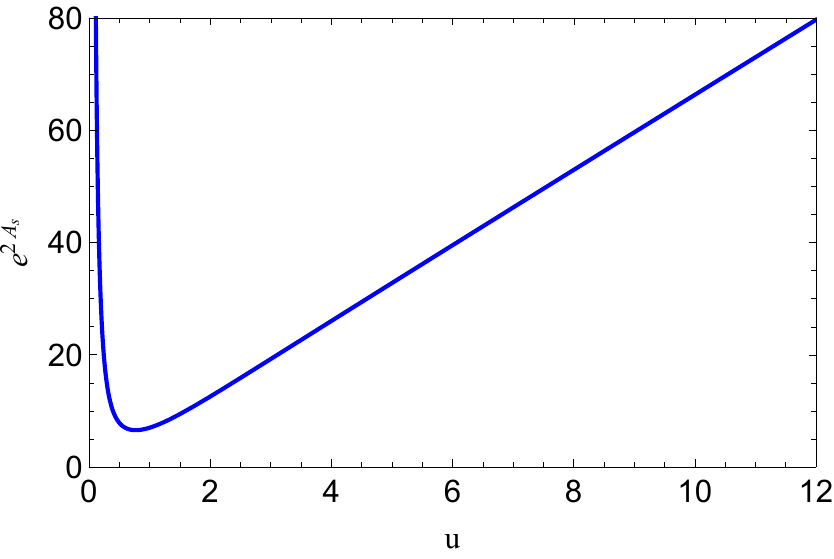}
\caption{
The exponential of the warp factor in string frame as a function of the dimensionless coordinate $u$.
}
\label{Fig:WarpFactString}
\end{figure}
\noindent

Once we have the warp factor, one can investigate if our space-time metric presents linear confinement. Here we follow the general criterion found in \cite{Kinar:1998vq}. where one considers the following family of metrics:
\begin{equation}
    ds^2 = -G_{00}(u) \, dt^2 + G_{x x} (u) \, dx^2 + G_{u u} (u) \, dz^2 + G_{x_T x_T}(u) \, dx_{T}^2 \, ,
\end{equation}
where $x$ are the $\mathbf{R}^{3}$, $u$ is the holographic coordinate and $x_{T}$ are the coordinates transverse to the corresponding five-dimensional space. In this analysis it is convenient to define the auxiliary function $f^{2}(u) = G_{00} \, G_{x x}$. Considering a rectangular Wilson loop in the 4d field theory, associated with quark anti-quark potential, the authors of  \cite{Kinar:1998vq} showed that confinement occurs if $f(u)$ has a global minimum at $u=u_0$ such that $f(u_{0})>0$.

In order to investigate our particular model, we plot the corresponding $f(u) = e^{2A(u)}$  in  Fig.~\ref{Fig:WarpFactString}. It is clear that the exponential of the warp factor in string frame has a global minimum  $f(u_0)>0$, where $u_0$ is the position of the minimum. Moreover, calculating the rectangular Wilson loops one obtains a linear quark antiquark potential  at large distances  $V(L) = \sigma L $ where $\sigma \propto f(u_0)$ is the corresponding string tension. Therefore, we conclude that our space-time metric presents linear confinement by the Wilson loop criterion.

At this point, it is worth pointing out that in the IR (large $u$), the exponential of the warp factor in the string frame diverges as $e^{2A_s}\sim u$. This behaviour shall be important below, when we solve the differential equation of the tachyon field.

\subsection{Spontaneous chiral symmetry breaking}
\label{Sec:XSB}

The inclusion of 4d mesons and chiral symmetry breaking in holographic QCD can be done as follows. The 4d currents associated with the global $SU(N_f)_L \times SU(N_f)_R$ symmetry (chiral symmetry) are mapped to 5d non-Abelian gauge fields while the quark mass operator associated with chiral symmetry breaking is mapped to a 5d scalar field (the tachyon). The 5d tachyon and gauge fields are described by a non-Abelian Higgs action proposed in \cite{Erlich:2005qh} and then generalized in \cite{Ballon-Bayona:2021ibm} to incorporate non-minimal dilaton couplings. In this work we only consider a non-minimal dilaton coupling in the tachyon potential term. The 5d action in the string frame takes the form
\begin{align}
S &= - \int d^5 x \sqrt{-g} \, e^{-\Phi}\,  \Big \{{\rm Tr} \Big ( |D_m X|^2 + f(\Phi)V(|X|)   \Big ) +\frac{1}{4 g_5^2} \,
{\rm Tr} \Big ( {F_{mn}^{(L)}}^2 +{F_{mn}^{(R)}}^2 \Big ) \Big \}  \,,  \label{5dmodel}
\end{align}
where $g_5^2=12\pi^2/N_c$ ($N_c$ number of colors) and the field strengths are given by
\begin{align}
F_{mn}^{(L/R)} &= \partial_m A_n^{(L/R)} - \partial_n A_m^{(L/R)} - i [ A_m^{(L/R)} , A_n^{(L/R)} ]  \\
D_m X &= \partial_m X - i A_m^{(L)} X + i X A_m^{(R)} \, .
\end{align}
\noindent
$X$, the tachyon field, represents a scalar field dual to the quark mass operator $\langle \bar{q} q \rangle$ and $V(X)$ the Higgs potential
\noindent
\begin{equation}
V(|X|) = m_X^2 |X|^2 + \lambda |X|^4 \,, 
\end{equation}
\noindent
where $m^2_X=-3$ is the mass term of the tachyon field and $\lambda$ a dimensionless coupling  that controls the minimum of the potential. The five-dimensional background has the form \eqref{Eq:metric} with the Einstein frame warp factor $A(u)$ replaced by the string frame warp factor $A_s(u)$ given in \eqref{Eq:stringwarpfactor}. The Einstein frame warp factor and dilaton field are given by \eqref{Eq:WarpFactorEinstein} and \eqref{Eq:DilatonSol} respectively. We remind the reader that we work in the quenched limit of holographic QCD so that the tachyonic and gauge fields do not backreact on the metric and dilaton field.   

In the following, we consider the case of two flavours in the dual field theory, i.e.  $N_f=2$. Then the model will describe the 5d gauge symmetry breaking $SU(2) \times SU(2) \to SU(2)$ which is the holographic dual of the 4d global symmetry breaking $SU(2) \times SU(2) \to SU(2)$ (chiral symmetry breaking) with 2 light flavours in QCD, as explained for instance in Refs.~\cite{Ballon-Bayona:2020qpq, Ballon-Bayona:2021ibm}. Assuming isospin symmetry, i.e. $m_u=m_d$ we take the ansatz $X(u)=\frac{1}{2}v(u)I_{2\times2}$ for the tachyon field and obtain the following differential equation 
\noindent
\begin{equation}\label{Eq:TachyonEq}
v''(u)+\left(3A_s'(u)-\Phi'(u)\right)v'(u)-e^{2A_s}f(\Phi)\left(m_X^2v-\frac{\lambda}{2}v^3\right)=0.
\end{equation}
\noindent
In the simplest case one may consider $f(\Phi)=1$ and $\lambda=0$. In this case the solution of the differential equation in the IR diverges in the form $v=C_0\,e^{u}$ \footnote{This solution is similar to the one obtained in  \cite{Gursoy:2007er} in the improved holographic QCD models (see Eq. 6.64).}. However, the task of extracting the condensate and quark mass becomes complicated due to the oscillatory behavior of the numerical solution close to the boundary, see Appendix \ref{Sec:DiverModel} for further analysis.

In the general case, $f(\Phi)$ is a non-trivial function of the dilaton field, while the parameter $\lambda$ is different from zero. Note also that Eq.~\eqref{Eq:TachyonEq} is invariant under the transformation $v\to \lambda^{-1/2} \,v$. This means that we can rewrite it in a form that the parameter $\lambda$ is hidden
\noindent
\begin{equation}\label{Eq:Tachyon}
v''(u)+\left(3A_s'(u)-\Phi'(u)\right)v'(u)-e^{2A_s}f(\Phi)\left(m_X^2v-\frac{1}{2}v^3\right)=0.
\end{equation}
\noindent
In the previous section we obtained the warp factor $A_s$ and the dilaton field $\Phi$, which are confined solutions of the Einstein-dilaton equations. The remaining challenge is to determine the unknown function $f(\Phi)$ for the model to guarantee spontaneous chiral symmetry breaking. In doing so, note that $e^{2A_s}\sim u$ in the IR, we realized that this term will ``induce'' a divergent solution for the tachyon in the IR. Note that the origin of this divergence is the fact that the actual model has an IR deformed AdS metric that has linear confinement. In order to attenuate this divergent behavior we build $f(\Phi)$ such that the product $e^{2A_s}f(\Phi)$ goes to zero in the IR. This condition shall guarantee a regular solution in this region and is analogous to what we have in the holographic soft wall model \cite{Ballon-Bayona:2020qpq, Ballon-Bayona:2021ibm}. This behavior also guarantees a non-oscillatory numerical solution close to the boundary, such that we do not find any problem extracting the quark mass and chiral condensate from the numerical solution. We consider the following ansatz for the non-minimal dilaton coupling
\noindent
\begin{equation}\label{Eq:f2}
f(\Phi)=\frac{1}{1+b_0\Phi+a_0\Phi^2}.
\end{equation}
\noindent
Note that i)  we have introduced two additional parameters $a_0$ and $b_0$, ii)  The asymptotic behaviour of this function is given by $f(\Phi)\to 1$ in the UV and $f(\Phi)\to 0$ in the IR. Later on, we will show that $a_0$ controls the presence of spontaneous chiral symmetry breaking in the model. In turn, $b_0$ will increase the value of the chiral condensate in the physical region, where the quark mass is positive, in order to avoid negative values. 
Furthermore, the function $f(\Phi)$ in the IR behaves like $ 1/u^4$ for $a_0>0$, see Eq.~\eqref{Eq:PHI_IR}, turning the dynamical source of spontaneous chiral symmetry breaking vanishing in the region where the strong confinement is present. Note that our constraint, a physical condition, are the Regge trajectories in this region. Therefore, one can think about $f(\Phi)$ separating the confinement (IR) and dynamical chiral symmetry breaking (UV) regions regulating the tachyon potential, see the action \eqref{Sec:XSB}, which seems physically reasonable as the latter is a phenomenon that also appears in theories where confinement is absent.

Now one may solve the differential equation in the asymptotic regions. Let us start the analysis close to the boundary, where we find from equation \eqref{Eq:Tachyon} that 
\noindent
\begin{equation}\label{Eq:TachyonUV}
v(u)=c_1\,u+c_3\,u^3+d_3\,u^3\,\ln{u}+\cdots \, ,
\end{equation}
\noindent
where the coefficient $d_3$ is given by
\noindent
\begin{equation}
\begin{split}
d_{3}=\,&\frac{c_1}{8}\left(2c_1^2\lambda+9\sqrt{10}\,\left(b_0-2\right)\right).
\end{split}
\end{equation}
\noindent
From the holographic dictionary we learned that the parameter $c_1$ is related to the quark mass in the 4d theory through the relation  $\Lambda c_1=\zeta\, m_q$, while the parameter $c_3$ is related to the chiral condensate $\Sigma$ through $\Lambda^{3}c_3=\Sigma/(2\zeta)$ \cite{Ballon-Bayona:2020qpq}. The parameter $\zeta=\sqrt{N_c}/(2\pi)$ was fixed with the expected results in the limit of large number of colors $N_c$ \cite{Cherman:2008eh}. The source coefficient $c_1$ and the VEV coefficient $c_3$ will be later read off  by matching the numerical solution to the asymptotic solution in the UV \eqref{Eq:TachyonUV}. Meanwhile, the asymptotic solution in the IR is given by 
\noindent
\begin{equation}
v=C_0+\frac{C_1}{u}+\frac{C_2}{u^2}+\frac{C_3}{u^3}+\cdots,
\end{equation}
\noindent
again, the coefficients were calculated solving \eqref{Eq:Tachyon} order-by-order in $u$
\noindent
\begin{equation}
\begin{split}
C_{1}=\,&0;\qquad\qquad C_{2}=0;\qquad\qquad C_{3}=\frac{2C_0}{81\,a_0}\exp\left(\frac43\Phi^{\infty}\right)(C_0^2\lambda-6).
\end{split}
\end{equation}
\noindent
Note that the function $f(\Phi)$ guarantees a regular behavior of the tachyon field in the IR.

\begin{figure}[ht!]
\centering
\includegraphics[width=7cm]{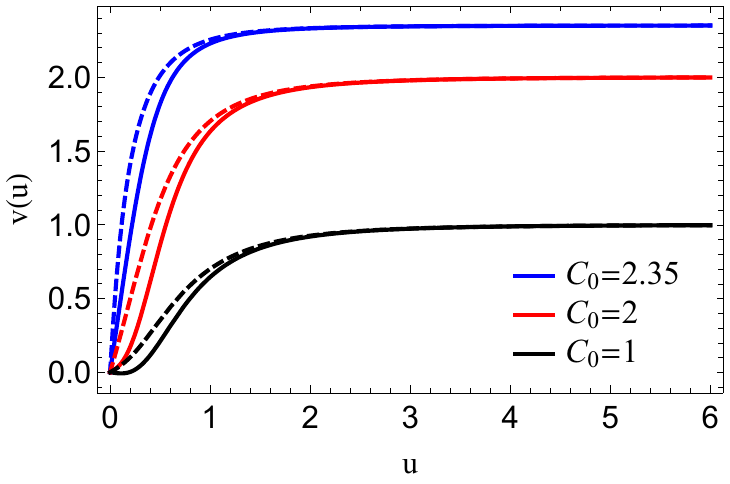}
\caption{
Profile of the tachyon field obtained in the model for $a_0=1$, selected values of $b_0$, $b_0=0$ (solid lines) and $b_0=1$ (dashed lines) and varying $C_0$ (different colors).
}
\label{Fig:Tachyon}
\end{figure}

Having obtained the asymptotic solutions, we can  use them as boundary conditions to solve the differential equation numerically. As the solution in the IR depends  only on one parameter, $C_0$, we solve the problem integrating numerically from the IR to the UV.\footnote{As a check of consistency we also integrated the differential equation from the UV to the IR obtaining the same results.} A plot of the tachyon profile, considering $a_0 =1$ and $b_0=0$, can be seen in Fig.~\ref{Fig:Tachyon}, where solid blue line represents the results for $C_0=2.35$, solid red line for $C_0=2$ and solid black line for $C_0=1$. Meanwhile, the corresponding results considering $a_0=1$ and $b_0=1$ are represented with dashed lines. Note that the effect of the parameter $b_0$ lies in the region close to the boundary, where we read off the chiral condensate, for that reason we expect that the value of the chiral condensate will be sensitive to the value of $b_0$.

\begin{figure}[ht!]
\centering
\includegraphics[width=7cm]{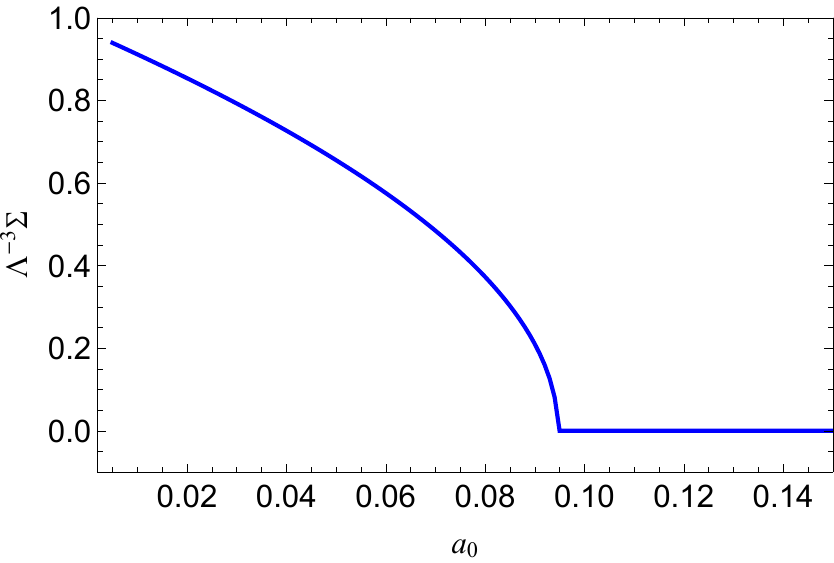}
\caption{
The dimensionless chiral condensate in the chiral limit as a function of $a_0$ for $b_0=1.7$.
}
\label{Fig:a0C0}
\end{figure}

Let us start by investigating the chiral limit, where $c_1=0$ (or $m_q=0$). In this case, we obtain a numerical relation between the chiral condensate, $\Sigma$, and the parameter $a_0$ as displayed in Fig.~\ref{Fig:a0C0}. As can be seen, there is a region where the chiral condensate is finite, characterizing the presence of spontaneous symmetry breaking, and a region where the chiral condensate is zero, characterizing the absence of spontaneous chiral symmetry breaking. The transition between these two regions occurs at $a_0^c\approx 0.0974$. From here on, we are going to fix the value of $a_0$ in the region of spontaneous chiral symmetry breaking.

\begin{figure}[ht!]
\centering
\includegraphics[width=7cm]{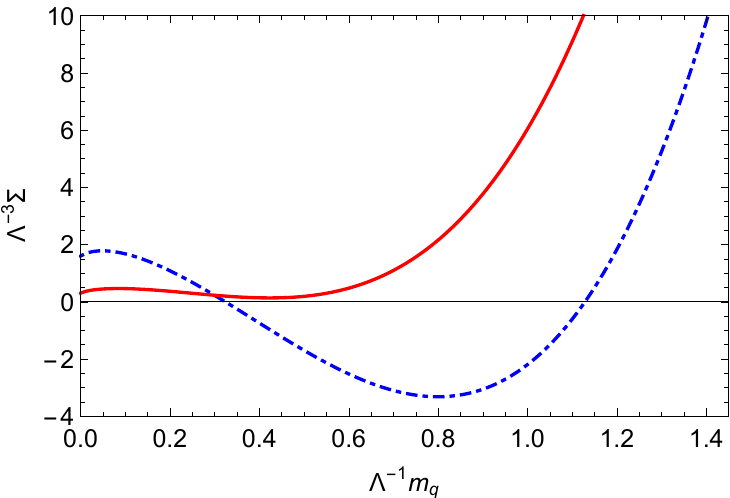}
\caption{
The dimensionless chiral condensate $\Lambda^{-3}\Sigma$ as a function of the dimensionless quark mass $\Lambda^{-1}m_q$ in the physical region, i.e, $m_q>0$. Blue dot-dashed line represents the results for $a_0=1$ and $b_0=0$, while red line were obtained for $a_0=0.02$ and $b_0=1.7$.
}
\label{Fig:c1c3}
\end{figure}

We now investigate the dependence of the chiral condensate on the quark mass. Let us consider two different values for $a_0$ and $b_0$. Our numerical results are displayed in Fig.~\ref{Fig:c1c3}, where blue dot-dashed line represents the results for $a_0=1$ and $b_0=0$, while the red line represents results for $a_0=0.02$ and $b_0=1.7$. As can be seen, the chiral condensate is nonzero in the chiral limit in both cases. Moreover, depending on the values of the parameters $a_0$ and $b_0$, the chiral condensate becomes negative (for $a_0=1$ and $b_0=0$) or positive (for $a_0=0.02$ and $b_0=1.7$) in the intermediate quark mass region. Then, it increases monotonically in the region of heavy quarks. The quark mass dependence of the chiral  condensate found in this work has a qualitatively similar behavior to the one obtained in references \cite{Iatrakis:2010jb, Ballon-Bayona:2021ibm}. Additional discussions, numerical details and more figures are given in Appendix \ref{Sec:AddNumerical} to complement the discussions of this section.

\section{Meson spectrum}
\label{Sec:Mesons}

In this section we calculate the spectrum for the vector, scalar, axial-vector and pseudoscalar mesons. It turns out that, excluding the vectorial sector, all mesons are highly sensitive to the tachyon field. The equations of motion describing these sectors may be obtained expanding the action \eqref{5dmodel} up to second order on the fields; for a detailed analysis see \cite{Ballon-Bayona:2020qpq, Ballon-Bayona:2021ibm} and  references therein. Here we write the relevant equations  and refer to those references for a detailed derivation. 

The scalar and pseudoscalar mesons in the dual field theory are mapped to the 5d scalar fields $S(u, \tilde x^{\mu})$ and $\pi^a(u, \tilde x^{\mu})$, with $a=1,2,3$ (adjoint representation of the $SU(2)$ isospin group). These fields arise as fluctuations of the tachyon field,
\noindent
\begin{equation}
\begin{split}
2X=\,&e^{2i \pi }(v+S) \, ,
\end{split}
\end{equation}
\noindent
where $\pi = \pi^a T^a$ and $T^a$ are the $SU(2)$ generators. In order to investigate the spectrum of the vector and axial-vector mesons, it is convenient to write the gauge fields $A^{(L/R)}_m$, appearing in the action \eqref{5dmodel}, in terms of a vectorial and axial gauge fields, $A^{(L/R)}_m=V_m\pm A_m$ where $V_m = V_{m,a} T^a$ and $A_m = A_{m,a} T^a$. The vectorial gauge symmetry allows us to fix $V_u=0$, while the residual axial gauge symmetry also allows us to fix $A_u=0$, see the discussion in \cite{Ballon-Bayona:2021ibm}. Thus, the fields have only four components $V_{\hat\mu}$ and $A_{\hat\mu}$. These vectors can further be decomposed into a transverse and longitudinal components, i.e., a Lorentz decomposition,
\noindent
\begin{equation}
\begin{split}
V_{\hat \mu, c} =\,& V_{\hat \mu,c}^{\perp} + \partial_{\hat \mu} \xi^c,\\
A_{\mu,c}=\,&A^{\perp}_{\mu,c}+\partial_{\mu}\varphi_c,
\end{split}
\end{equation}
\noindent
the indices $(\hat \mu, \hat \nu)$ correspond to coordinates in four-dimensional flat spacetime, while, the index $c=(1,2,3)$ corresponds to the adjoint representation of the $SU(2)$ isospin group. One can set $\xi^c=0$, because it does not have dynamics. Thus, the vector and axial-vector mesons are described by the fields $V^{\perp}_{\hat \mu,c}$ and $A^{\perp}_{\hat \mu,c}$ while the pseudoscalar mesons are described by the field $\pi_c$ coupled to the longitudinal component of the axial-vector field $\varphi_c$. Following the steps described in \cite{Ballon-Bayona:2020qpq, Ballon-Bayona:2021ibm} we find the following set of differential equations  
\noindent
\begin{align}
& \Big [ \partial_u + 3 A_s' - \Phi' \Big ] \partial_u S 
+ \Box S - e^{2 A_s}f(\Phi) \frac{d^2 U}{dv^2} S = 0  \quad 
({\rm scalar \, sector}) \, , \label{Seq} \\ 
& \Big [ \partial_u + A_s' - \Phi' \Big ] \partial_u V^{\hat \mu,c}_{\perp} + \Box   V^{\hat \mu,c}_{\perp}   
 =0 \quad ({\rm vectorial \, sector}) \, , \label{Veq}\\
 & \Big [ \partial_u + A_s' - \Phi' \Big ] \partial_u A^{\hat \mu,c}_{\perp} + \Box A^{\hat \mu,c}_{\perp} -  \beta A^{\hat \mu,c}_{\perp} = 0 \quad ({\rm axial \, sector}) \, , \label{Aeq}  \\
&\Big [ \partial_u + A_s' - \Phi' \Big ] \partial_u  \varphi^c +   \beta (  \pi^c  -  \varphi^c )  = 0 \,  \quad (\text{pseudoscalar sector})\label{varphieq}, \\
& - \partial_u \Box \varphi^c + \beta  \, \partial_u \pi^c  = 0 
\quad (\text{pseudoscalar sector}) \,, \label{pieq}
\end{align}
\noindent
where $\beta=g_5^2\,v^2\,e^{2A_s}$.

Before describing our results for the meson spectrum we present in Table \ref{tab:Parameters} the set of final values for the model parameters we are going to use in the forthcoming analysis. The parameter $\Lambda$ is fixed in order to reproduce the mass of the fundamental state in the vectorial sector, i.e. the $\rho(770)$ meson. The parameter $a_0$ is fixed considering the mass of the first excited state in the scalar sector, the $f_0(980)$. meson. The parameters $\lambda$ and $m_q$ in model A are fixed in order to reproduce the experimental value for the masses of the first two pseudoscalar states, i.e. $\pi^{+}$ and $\pi(1300)$ while  the parameters in model B are fixed in order to reproduce the experimental value for the mass and decay constant of the first pseudoscalar state, i.e. $\pi^{+}$.
%%%%%%%%%%%%%%%%%%%%%
\begin{table}[ht]
\centering
\begin{tabular}{l |c|c}
\hline 
\hline
 Parameter & Model A & Model B  \\
\hline 
 $a_0$ & $0.02$  & $0.02$  \\
 $b_0$ & $1.7$  & $1.7$  \\
 $\Lambda$ & $369\,\text{MeV}$  &  $369\,\text{MeV}$  \\
$\lambda$ & $27$  & $2.4$\\
$m_q$ & $5.49\,\text{MeV}$  & $5.49$\,\text{MeV}\\
\hline
\hline
\end{tabular}
\caption{
Final values for the parameters in models A and B
}
\label{tab:Parameters}
\end{table}
%%%%%%%%%%%%%%%%%%%%%

\subsection{Vector mesons}

Let us start by calculating the spectrum of the vector mesons, which is decoupled from the other sectors. In order to obtain the spectrum of vector mesons  from Eq.~\eqref{Veq}, we first take the 4d Fourier transform  $V_{\mu}^{\perp}(\tilde x^{\mu},u)\to V_{\mu}^{\perp}(\tilde k^{\mu},u)$, so that  $\square\to - \tilde k^2$. Then one takes the ansatz $V_{\mu}^{\perp} = \eta_{\mu} v(\tilde k, u)$, with $\eta_{\mu}$ a (transverse) polarisation vector, i.e.  $\eta_\mu \tilde k^{\mu}=0$,  so that we arrive at a second order differential equation for $v(\tilde k, u)$. The eigenvalue problem corresponds to setting $\tilde k^2 \to - m_{V_n}^2/\Lambda^2$ \footnote{The vector components $\tilde k_{\mu}$ and $\tilde x_{\mu}$ are dimensionless.} and $v(k,u) \to v_n(u)$. One can solve the eigenvalue problem directly or write the differential equation in the Schr\"odinger-like form through the ansatz for a normalizable mode $v_n(u)=  e^{-B_{V}} \psi_{v_n}(u)$ where $2B_{V}=A_s-\Phi$. Then, we calculate the spectrum solving the Schr\"odinger-like differential equation
\noindent
\begin{equation}\label{Eq:SchrodingerVector}
-\partial^2_u\,\psi_{v_n} +V_{V}\,\psi_{v_n} =\frac{m_{V_n}^2}{\Lambda^2}\, \psi_{v_n},
\end{equation}
\noindent
where the potential is given by
\begin{equation}\label{Eq:VectorPot}
V_{V}=\left(\partial_u B_{V}\right)^2+\partial^2_{u} B_{V}.
\end{equation}
\noindent
We obtain numerically the eigenvalues using a shooting method. Then, we fix the parameter $\Lambda=369$ MeV by comparing the first eigenvalue with the  experimental result for the mass of the $\rho(770)$ meson, extracted from particle data group (PDG) \cite{Workman:2022ynf}. Having fixed $\Lambda$ we calculate the masses of the excited states; the results are displayed in Table \ref{Tab:VectorMesons}. As can be seen from the table, our results are in quantitative agreement with the experimental data. 

%%%%%%%%%%%%%%%%%%%%%
\begin{table}[tbp]
\centering
\begin{tabular}{l |c|l}
\hline 
\hline
 $n$ & Model &
$\rho$ experimental \cite{Workman:2022ynf} \\
\hline 
 $0$ &  776 & $776\pm 1$  \\
 $1$ & 1170  & $1282\pm 37$  \\
 $2$ & 1468  & $1465\pm 25$ \\
 $3$ & 1719  & $1720\pm 20$  \\
 $4$ & 1939  & 1840 to 1951   \\
 $5$ & 2136  & 1910 to 2322 \\
 \hline\hline
\end{tabular}
\caption{
Masses of vector mesons (in MeV), compared against the experimental data from PDG \cite{Workman:2022ynf}. The experimental result for the first excited state, $n=1$, was taken from \cite{OBELIX:1997zla} The results for the states $n=4$ and $n=5$ were roughly estimated from the results for $\rho(1900)$  and $\rho(2150)$ given in \cite{Workman:2022ynf}. We have omitted in this table the state $\rho(1570)$ because according to \cite{Workman:2022ynf} it may be a OZI violating decay mode of the  $\rho(1700)$, the latter appears in the table as the $n=3$ state.}
\label{Tab:VectorMesons}
\end{table}
%%%%%%%%%%%%%%%%%%%%%

\subsection{Scalar mesons}

In order to obtain the spectrum of  scalar mesons we first take the 4d Fourier transform of equation \eqref{Seq}, i.e. $S(\tilde x^{\mu},u) \to S(\tilde k^{\mu},u)$ and $\Box \to -\tilde k^2$.  The eigenvalue problem corresponds to setting $\tilde k^2 \to -m_{S_n}^2/\Lambda^2$ and $S(k,u) \to S_n(u)$.  It can be solved considering the ansatz  $S_n(u)=e^{-B_S}\psi_{s_{n}}(u)$, where $2 B_S = 3 A_s - \Phi$, for the normalizable mode so that the differential equation takes the Schr\"odinger-like form
\noindent
\begin{equation}\label{Eq:SchroScalarEq}
-\partial_u^2 \psi_{s_{n}}+V_S\,\psi_{s_{n}}=\frac{m_{S_n}^2}{\Lambda^2}\,\psi_{s_{n}},
\end{equation}
\noindent
where the potential is given by
\noindent
\begin{equation}\label{Eq:Schro/potScalar1}
V_S=(\partial_u B_S)^2+\partial_u^2B_S
+e^{2A_s}f(\Phi)\left(m_X^{2}+\frac32\,\lambda\,v^2\right).
\end{equation}
\noindent
The relevance of the function $f(\Phi)$ may be highlighted here. Note that close to the boundary, i.e., UV region, $f(\Phi)\to 1$, the leading term is the warp factor, i.e., $B_S\sim 3A_s/2$, with $A_s=-\ln{u}$, so the potential goes like $V_S\sim \frac{3}{4u^2}$. Meanwhile, in the IR region, $f(\Phi)\to 0$, the leading term is the dilaton, i.e., $B_S\sim -\Phi/2$, with $\Phi=3u^2$/2, and the potential goes like $V_S\sim 9u^2/4$. This means that we will obtain asymptotically linear Regge trajectories.

Before evaluating the spectrum of scalar mesons for the non-trivial tachyon solution we take a moment to analyze the instability of
the trivial tachyon solution, i.e., $v=0$. In this case the Schr\"odinger potential in the scalar sector reduces to
\noindent
\begin{equation}\label{Eq:Schro/potScalar2}
V_S=(\partial_u B_S)^2+\partial_u^2B_S
+\frac{e^{2A_s}}{1+b_0\Phi+a_0\,\Phi^2}m_X^{2}.
\end{equation}
\noindent

\begin{figure}[ht!]
\centering
\includegraphics[width=7cm]{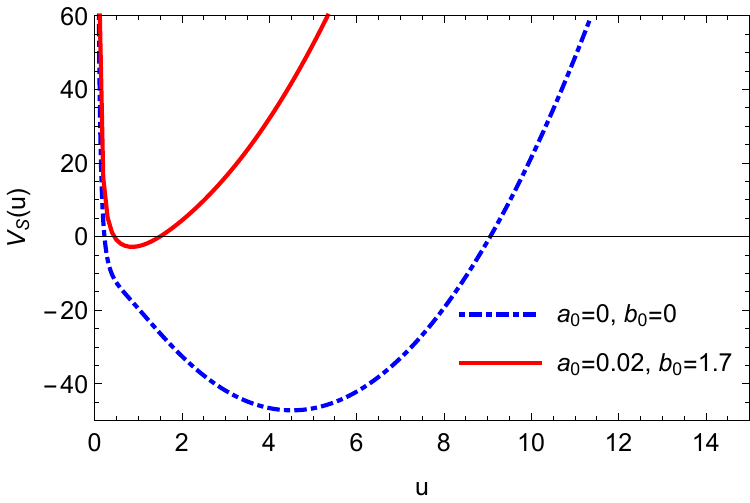}
\caption{
The potential of the Schr\"odinger-like equation for the trivial tachyon, with blue dot-dashed line, and non-trivial solution in the chiral limit, with red line, for $\lambda=27$.
}
\label{Fig:PotScalar}
\end{figure}
\noindent
Note that the potential depends on the parameters $b_0$ and $a_0$. A plot of this potential for $a_0=0$ and $b_0=0$ is displayed in Fig.~\ref{Fig:PotScalar} with blue dot-dashed line. We can see from the figure that the global minimum of the potential is negative, supporting states with negative mass-squared, $m_{S_n}^2<0$, i.e., instabilities. In turn, when we turn on the tachyon field and increase the parameters $a_0$ and $b_0$ to $a_0=0.02$ and $b_0=1.7$, the negative part of the potential well decreases, eliminating the instabilities, $m_{S_n}^2>0$, see the solid red line in Fig.~\ref{Fig:PotScalar}. 

\begin{figure}[ht!]
\centering
\includegraphics[width=7cm,height=4.5cm]{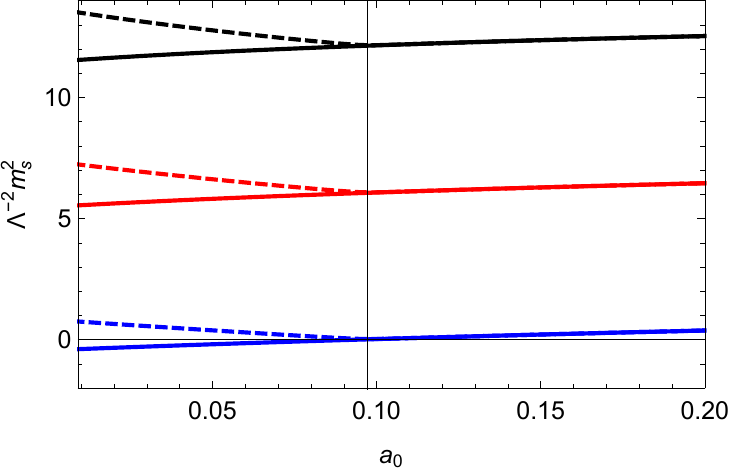}
\caption{
The mass-squared of scalar mesons in the chiral limit as a function of $a_0$. Solid lines represent the results for the trivial tachyon with $b_0=0$, while dashed lines represent the non-trivial tachyon, with $b_0=1.7$. These results were obtained for $\lambda=27$.
}
\label{Fig:Massv0}
\end{figure}

For observing instabilities in the spectrum, we plot the mass-squared of scalar mesons in the chiral limit, i.e. $c_1=0$, as a function of $a_0$ in Fig.~\ref{Fig:Massv0}. In this figure, solid lines represent the spectrum obtained considering the trivial tachyon. We observe that the instabilities disappear increasing the value of $a_0$ up to some critical value, which is around $a_{0}^c\approx 0.0974$. These results can be interpreted as a transition from an unstable to stable background carried out by the parameter $a_0$. Let us now  turn our attention to the spectrum obtained considering a non-trivial tachyon. The mass-squared as a function of the parameter $a_0$ is displayed in Fig.~\ref{Fig:Massv0} with dashed lines. As can be seen from the figure, in the region of small $a_0$ there are no instabilities. The mass-squared decreases with the increasing of $a_0$ up to the critical value $a_0^c\approx 0.0974$. Then, for $a_0> a_{0}^c$, the mass-squared increases for increasing $a_0$ . Note that the mass-squared becomes degenerate in the region for $a_0>a_{0}^c$ in the sense that we get the same mass for the trivial and non-trivial tachyon.

\begin{figure}[ht!]
\centering
\includegraphics[width=7cm,height=4.5cm]{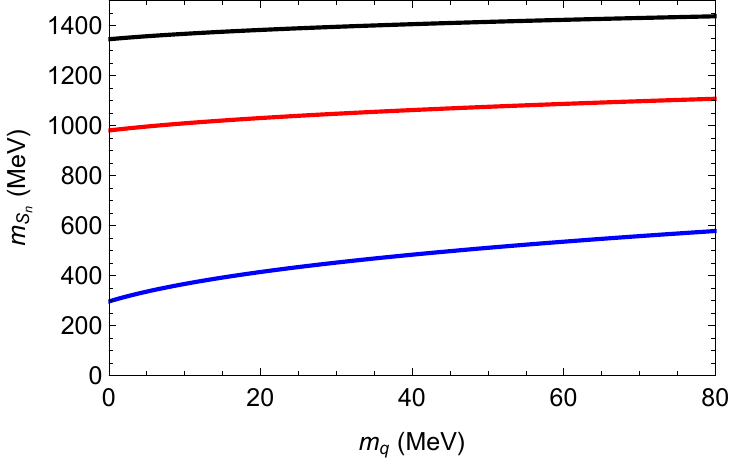}
\caption{
The mass of the scalar mesons as a function of the quark mass for $a_0=0.02$, $b_0=1.7$ and $\lambda=27$.
}
\label{Fig:ScaMassmq}
\end{figure}

In Fig. \ref{Fig:ScaMassmq} we plot the mass of the scalar mesons for the non-trivial tachyon as a function of the quark mass. The model parameters were fixed as in Table \ref{tab:Parameters} except for the parameter $c_1$ that varies with the quark mass.  As can be seen from the figure, the mass of the scalar mesons remains finite in the chiral limit. Then, it increases with the quark mass. Note that the ground state is more sensitive to the variation of the quark mass. 

Finally, we show our final results for the spectrum of scalar mesons using the parameters displayed in Table~\ref{tab:Parameters}. Our results are displayed in Table \ref{Tab:ScalarMesons}, compared against the corresponding experimental results reported in Ref.~\cite{Workman:2022ynf}.

%%%%%%%%%%%%%%%%%%%%%
\begin{table}[tbp]
\centering
\begin{tabular}{l |c| c| l}
\hline 
\hline
 $n$ & Model A &  Model B &
$f_0$ experimental \cite{Workman:2022ynf} \\
\hline 
 $0$ & 338  & 310  &  400 to 800 \\
 $1$ & 997  & 985 & $990\pm 20$  \\
 $2$ & 1358 & 1349 & 1200 to 1500  \\
 $3$ & 1637 & 1630 & $1506\pm 6$ \\
 $4$ & 1873 & 1867 & $1704 \pm 12$  \\
 $5$ & 2081 & 2075 & $1992\pm 16$  \\
 $6$ & 2269 & 2265 & $2095^{+17}_{-19}$ \\
 \hline\hline
\end{tabular}
\caption{
Masses of the scalar mesons (in MeV), compared against the experimental data from PDG \cite{Workman:2022ynf}. These results were obtained for the set of parameters displayed in Table \ref{tab:Parameters}.
}
\label{Tab:ScalarMesons}
\end{table}
%%%%%%%%%%%%%%%%%%%%%

\subsection{Axial-vector mesons}

In order to calculate the spectrum of the axial-vector mesons, we first apply the 4d Fourier transform to the field $A^{\mu}_{\perp}(\tilde x^{\mu},u)\to A^{\mu}_{\perp}(\tilde k^{\mu},u)$ and $\square\to - \tilde k^2$ in Eq.~\eqref{Aeq}.  Then one takes the ansatz $A_{\mu}^{\perp} = \eta_{\mu} a(\tilde k, u)$, with $\eta_{\mu}$ a (transverse) polarisation vector, so that we arrive at a second order differential equation for $a(\tilde k, u)$. The eigenvalue problem corresponds to setting $\tilde k^2 \to - m_{A_n}^2/\Lambda^2$ and $a(k,u) \to a_n(u)$. Then, we redefine the normalizable mode as $a_n=  e^{-B_{A}} \psi_{a_n}$, where $2B_{A}=A_s-\Phi$ and we obtain the Schr\"odinger-like differntial equation
\noindent
\begin{equation}\label{Eq:AVSchroEq}
-\partial^2_u \psi_{a_n}+V_{A}\,\psi_{a_n}= \frac{m_{A_n}^2}{\Lambda^2} \,\psi_{a_n},
\end{equation}
\noindent
where the potential is given by
\begin{equation}\label{Eq:AVPotential}
V_{A}=\left(\partial_u B_{A}\right)^2+\partial^2_{u} B_{A}
+\beta,
\end{equation}
\noindent

\begin{figure}[ht!]
\centering
\includegraphics[width=7cm,height=4.5cm]{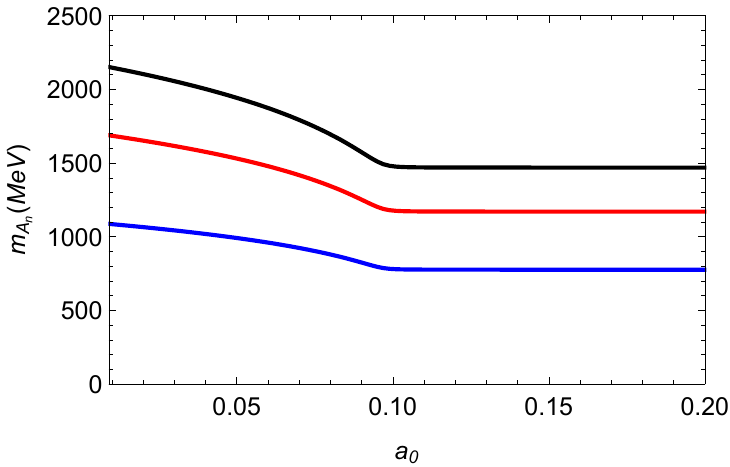}\hfill
\includegraphics[width=7cm,height=4.5cm]{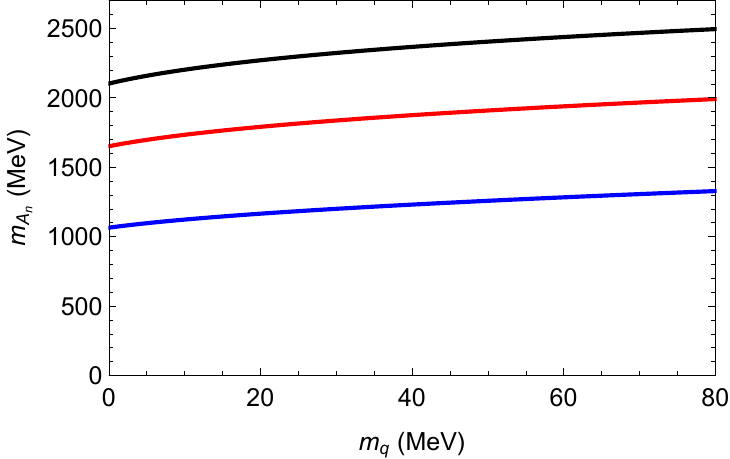}
\caption{
Left panel: The mass of the axial-vector mesons as a function of $a_0$ in the chiral limit for $b_0=1.7$ and $\lambda=27$. Right panel: The mass of the axial-vector mesons as a function of the quark mass for $a_0=0.02$, $b_0=1.7$ and $\lambda=27$.
}
\label{Fig:AVMassmq}
\end{figure}

We solve numerically the eigenvalue problem using a shooting method. First, we investigate how the massese of the axial-vector mesons change with the increasing of the parameter $a_0$ in the chiral limit. This is shown on the left panel of Fig.~\ref{Fig:AVMassmq}. As can be seen from the plot, the masses are finite in the limit of zero $a_0$. Then, they decrease up to $a_0=a_{0_c}$, lying approximately constant in the region $a_0> a_{0_c}$. Next, fixing the value of $a_0$ at $a_0=0.02$, we analyze the evolution of the masses as a function of the quark mass. This is displayed on the right panel of Fig.~\ref{Fig:AVMassmq}. As can be seen from the plot, the masses are finite in the chiral limit and then increase as the quark mass grows. Finally, using the parameters of Table~\ref{tab:Parameters} we calculate the spectrum and compare with experimental data. This is  displayed in Table \ref{Tab:AVMesons}.

%%%%%%%%%%%%%%%%%%%%%
\begin{table}[tbp]
\centering
\begin{tabular}{l |c|c|l}
\hline 
\hline
 $n$ & Model A & Model B &
$a_1$ experimental \cite{Workman:2022ynf} \\
\hline 
 $0$ & 1099  & 1602  &  $1230\pm40$  \\
 $1$ & 1700  & 2535  & $1655\pm16$ \\
 $2$ & 2163  & 3287  & $1930^{+30}_{-70}$ \\
 $3$ & 2548  & 3939  & $2096\pm138$\\
 $4$ & 2881  & 4522  & $2270^{+55}_{-40}$\\
 $5$ & 3177  & 5053 &  \\
 \hline\hline
\end{tabular}
\caption{
Masses of the axial-vector mesons (in MeV), compared against the experimental data from PDG \cite{Workman:2022ynf} for the states $n=0$ and $n=1$ and the experimental data from \cite{Zyla:2020zbs} for the states $n=2$, $n=3$ and $n=4$.
These results were obtained for the set of parameters displayed in Table \ref{tab:Parameters}.
}
\label{Tab:AVMesons}
\end{table}
%%%%%%%%%%%%%%%%%%%%%

\subsection{Pseudoscalar mesons}

\begin{figure}[ht!]
\centering
\includegraphics[width=7cm,height=4.5cm]{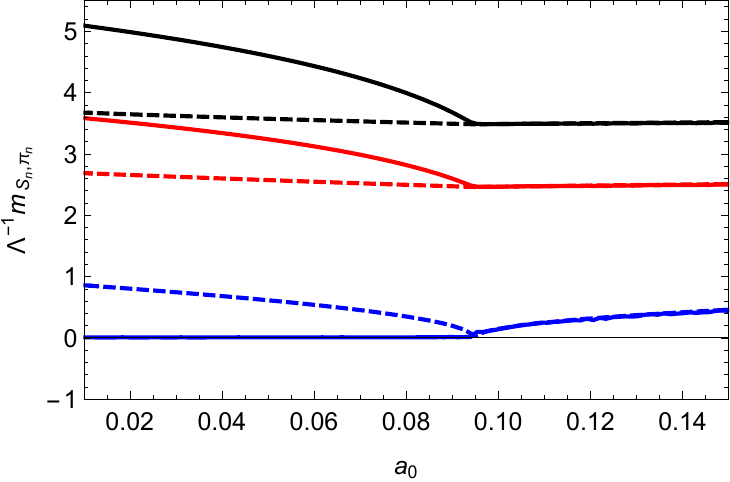}
\caption{
Masses of the scalar mesons (dashed lines), and pseudoscalar mesons (solid lines) as a function of the parameter $a_0$, $b_0=1.7$  and $\lambda=27$  in the chiral limit.
}
\label{Fig:MassScalarPseudoScalar}
\end{figure}

We now calculate the spectrum of the pseudoscalar mesons in the dual field theory. It is worth mentioning that we follow a different approach in this section due to the nature of the differential equations. Transforming into a Schrodinger-like differential equation does not bring any advantage in relation to a direct solution of the coupled differential equations, see for instance the analysis done in \cite{Ballon-Bayona:2020qpq}. Applying the 4d Fourier transform to the fields $\pi(\tilde x^{\mu},u)\to \pi(\tilde k^{\mu},u)$, $\varphi(\tilde x^{\mu},u)\to \varphi(\tilde k^{\mu},u)$ and we also have $\square\to - \tilde k^2$. The eigenvalue problem consists this time in solving  the coupled differential equations \eqref{varphieq} and \eqref{pieq} setting $\tilde k^2 \to - m_{\pi_n}^2$, $\pi(\tilde k ,u) \to \pi_n(u)$ and $\varphi(\tilde k, u) \to \varphi_n(u)$. The coupled differential equations become
\noindent
\begin{align}
\Big [ \partial_u + A_s' - \Phi' \Big ] \partial_u  \varphi_n +   \beta (  \pi_n  -  \varphi_n)  =\,& 0, \\
- m_{\pi_n}^2 \partial_u \varphi_n + \beta  \, \partial_u \pi_n  =\,& 0.
\end{align}
\noindent
We solve these coupled differential equations numerically using a shooting method. In order to compare the results against those obtained for the scalar sector, in the chiral limit, we implemented an overlap of the results in Fig.~\ref{Fig:MassScalarPseudoScalar}, where dashed lines represent the results for the scalar mesons, while solid lines represent the results for the pseudoscalar mesons. As can be seen from the figure, in the region of interest, i.e., $a_0<a_{0_c}$, the mass of the ground state in the scalar sector is finite, while the mass of the ground state of the pseudoscalar mesons is zero, i.e., the pion becomes a Goldstone boson in the chiral limit. In turn, the mass of the scalar resonances decreases faster than the mass of the pseudoscalar resonances for $a_0< a_{0_c}$. Interestingly, in the regime $a_0> a_{0_c}$ the scalar and pseudoscalar mesons become degenerate signalling the restoration of chiral symmetry (chiral limit).  

\begin{figure}[ht!]
\centering
\includegraphics[width=7cm]{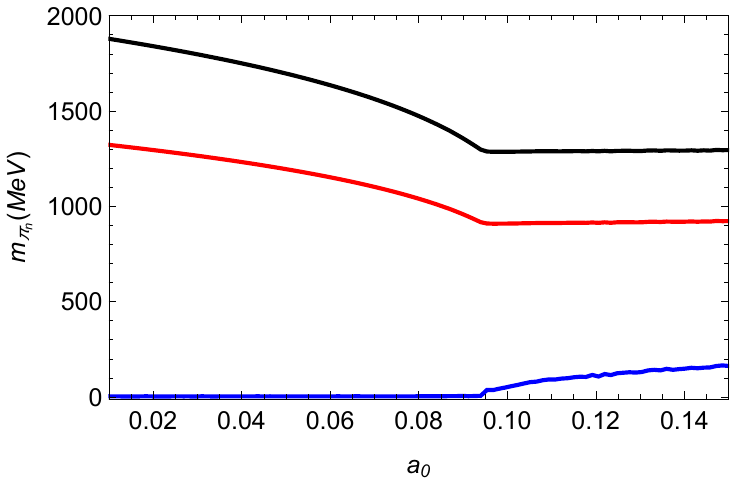}\hfill
\includegraphics[width=7.5cm]{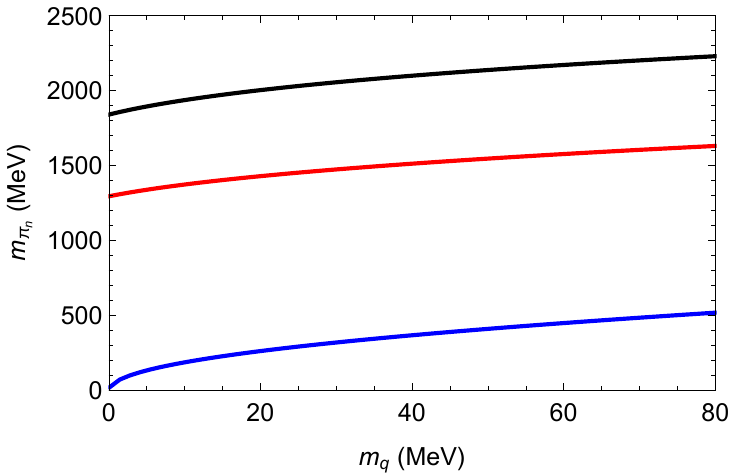}
\caption{
Left panel: The mass of the pseudoscalar mesons as a function of $a_0$ in the chiral limit, for $\lambda=27$ and $b_0=1.7$. Right panel: The mass of the pseudoscalar mesons as a function of the quark mass $m_q$, for $\lambda=27$, $a_0=0.02$ and $b_0=1.7$.
}
\label{Fig:mqmpi}
\end{figure}

We also calculate the masses of pseudoscalar mesons as a function of $a_0$ in the chiral limit, see the left panel of Fig.~\ref{Fig:mqmpi}. As expected, the ground state has zero mass in the region for $a_0< a_{0_c}$. Then, the mass increases for $a_0> a_{0_c}$. Next, we fix $a_0=0.02$ and analyze the masses as a function of the quark mass, see the right panel of Fig. \ref{Fig:mqmpi}. As can be seen from the plot, the mass of the lightest pseudoscalar state goes to zero in the chiral limit so that this state is identified as the pion. We also note that the masses of pseudoscalar mesons are less sensitive to the quark mass in the region of heavy-quarks.

We finally present our results for the masses of pseudoscalar mesons using the parameters of Table \ref{tab:Parameters}. As described at the beginning of this section, the parameters $\lambda$ and $m_q$ in model A were fixed using the masses of the first two pseudoscalar states, i.e. $\pi^{+}$ and $\pi(1300)$ while  the parameters in model B were fixed using the mass and decay constant of the first pseudoscalar state, i.e. $\pi^{+}$. Our final results for the pseudoscalar meson masses, compared against the experimental results, are displayed in Table \ref{Taba:Pseudoscalar}. 

%{\Blue [Comment] - O PDG \'e citado em duas refer\^encias: \cite{Zyla:2020zbs} e \cite{Tanabashi:2018oca} . Uma \'e de 2018 e a outra de 2020. Minha sugest\~ao \'e manter apenas a refer\^encia do PDG mais recente, que \'e de 2022 e inclu\'i no .bib : \cite{Workman:2022ynf}.}

%%%%%%%%%%%%%%%%%%%%%
\begin{table}[ht]
\centering
\begin{tabular}{l |c|c|l}
\hline 
\hline
 $n$ & Model A & Model B & $\pi$ experimental \cite{Workman:2022ynf} \\
\hline 
\hline 
 $0$ &  140  & 138  & $139.6\pm0.0002$  \\
 $1$ &  1342 & 2012 & $1300\pm 100$  \\
 $2$ & 1899  & 2877 & $1810^{+9}_{-11}$ \\
 $3$ & 2334  & 3588 &  \\
\hline\hline
\end{tabular}
\caption{
The mass of the pseudoscalar mesons (in MeV), compared against the experimental data \cite{Workman:2022ynf}. These results were obtained for the set of parameters displayed in Table \ref{tab:Parameters}.
}
\label{Taba:Pseudoscalar}
\end{table}
%%%%%%%%%%%%%%%%%%%%%

\subsection{Regge trajectories}

\begin{figure}[ht!]
\centering
\includegraphics[width=7cm]{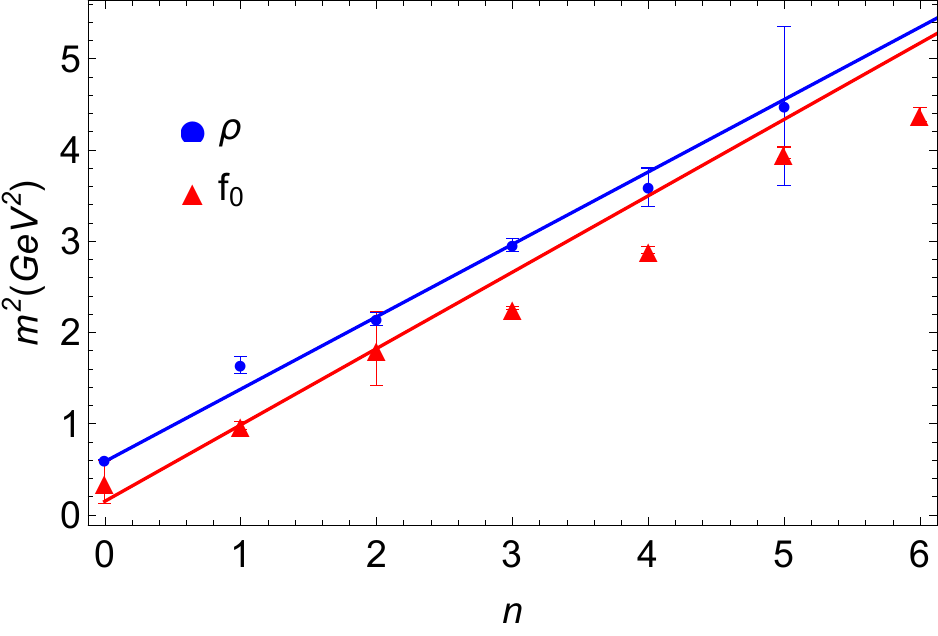}\hfill 
\includegraphics[width=7cm]{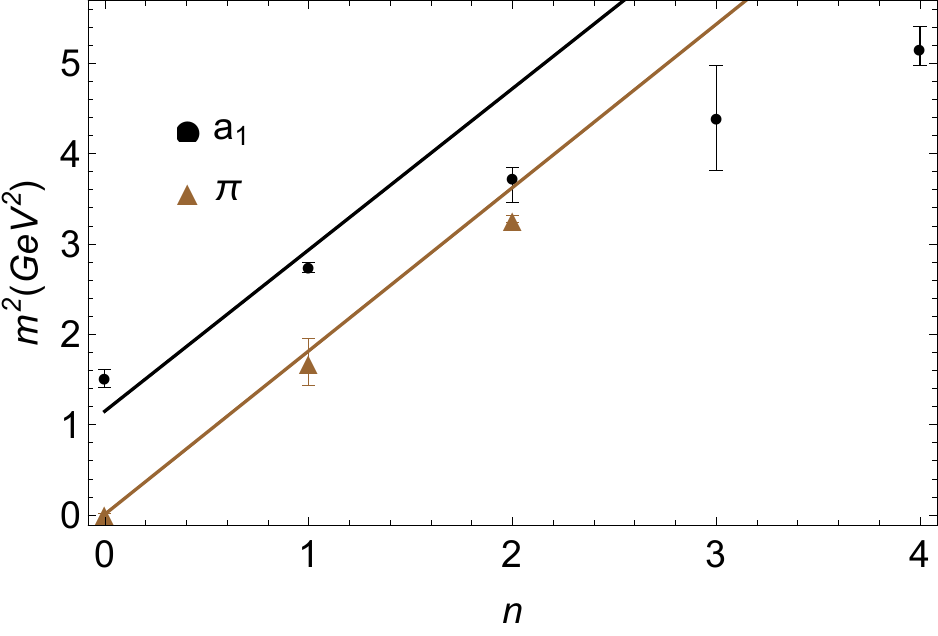}
\caption{
Linear Regge trajectories in model A compared against the experimental result for the mass-squared of mesons as a function of {the radial excitation} $n$. Left panel: vector meson spectrum obtained from Eq.(\ref{vec1}) (blue line), scalar meson spectrum obtained from Eq.(\ref{scal1}) (red line). Experimental data correspond to: rho meson family (dots), $f_0$ family (triangles). Right panel: axial-vector meson spectrum obtained from Eq.(\ref{axi1}) (black line), pseudoscalar meson spectrum obtained from Eq.(\ref{pseud1}) (brown line). Experimental data correspond to: $a_1$ meson family (dots), pion family (triangles). } 
\label{Fig:MassSquare}
\end{figure}

To end this section we use our results for the meson masses presented in tables \ref{Tab:VectorMesons}, \ref{Tab:ScalarMesons}, \ref{Tab:AVMesons} and \ref{Taba:Pseudoscalar} to perform linear fits for the mass-squared as a function of the radial excitation number $n$. We find the following approximately linear Regge trajectories:
\noindent
\begin{align}
    m_{V_n}^2=\,&\left(0.794\, n+0.583\right)\,\text{GeV}^2 \label{vec1}\\
    m_{S_n}^2=\,&\left(0.837\, n+0.148\right)\,\text{GeV}^2 \label{scal1}\\
    m_{a_n}^2=\,&\left(1.79\, n+1.15\right)\,\text{GeV}^2 \label{axi1}\\
    m_{\pi_n}^2=\,&\left(1.81\, n+0.00563\right)\,\text{GeV}^2 \label{pseud1}
\end{align}
\noindent
In Fig.~\ref{Fig:MassSquare} we compare our linear fits against the experimental results for the mass-squared for the vector, scalar, axial-vector and pseudoscalar mesons as a function of the radial excitation number.

We note from our results that, for the parameters considered in model A,  the slope for the vector and scalar Regge trajectories are half of the slope of the axial-vector and pseudoscalar Regge trajectories.  We performed a WKB analysis following  \cite{Gursoy:2007er}. For the vector and scalar mesons, we obtained the behavior of the squared mass $m^2_{V_n, S_n}=6\Lambda^2\,n$, from which we read off the slope, $0.817$ GeV$^2$, which is close to the value obtained in our fit. However, for the axial-vector the WKB analysis is more subtle because the tachyon field contributes in a nontrivial way to the asymptotic behavior of the potential \eqref{Eq:AVPotential}. A similar situation occurs in the pseudoscalar sector.  This means that the slopes appearing in equations \eqref{axi1} and \eqref{pseud1} are carrying some information about chiral symmetry breaking in addition to infrared confinement, while the slopes of equations \eqref{vec1} and \eqref{scal1} are only due to confinement.

We also compared quantitatively the results for model A against the experimental results calculating the $\chi^2$ test. For the vector mesons we obtained $\chi^2_{V_n}=0.81$, for the scalar mesons $\chi^2_{S_n}=0.99$, for the axial-vector mesons $\chi^2_{a_n}=0.83$, and for the pseudoscalar mesons $\chi^2_{\pi_n}=1$.

\section{Decay constants}
\label{Sec:Decay}

In the following, we use the prescription developed in Refs.~\cite{Ballon-Bayona:2020qpq, Ballon-Bayona:2021ibm}, see also references therein, where the task of calculating decays constants was reduced to the calculation of normalization constants of the wave functions obtained solving the eigenvalue problems.

\subsection{Vector mesons}

Let us start by calculating the decay constants for the vector mesons. The wave functions, which are solutions of the eigenvalue problem, are not normalized. Thus, there are normalization constants that can be obtained using the normalization condition
\noindent
\begin{equation}
\int\,du\, \psi_{v_m}(u)\psi_{v_n}(u)=\int du\, e^{A_s-\Phi}v_m(u)\,v_n(u)=\delta_{mn}, \label{Eq:Normvec}
\end{equation}
\noindent
where $A_s$ is the warp factor in string frame, $\Phi$ the dilaton field, $\psi_{v_n}$ is the solution of the Schrödinger-like equation \eqref{Eq:SchrodingerVector} and $v_n$ is the normalizable solution related to $\psi_{v_n}$ through $v_n(u)=\,e^{-B_{V}}\psi_{v_{n}}(u)$. As showed in \cite{Ballon-Bayona:2021ibm}, the vector meson decay constants reduce to the following formula:
\noindent
\begin{equation}
F_{v_n}=\lim_{\epsilon\to 0}\frac{e^{A_s-\Phi}}{g_5}\,\partial_z v_n\bigg{|}_{z=\epsilon}=\frac{2}{g_5}N_{v_n},
\end{equation}
\noindent
where $N_{v_n}$ is the normalisation constant, which is the coefficient of the normalizable solution in the expansion of the vector field close to the boundary, i.e.  $v_n(u)= N_{v_n} u^2 + \dots $. We calculate numerically the normalization constants and find the vector meson decay constants. Our results are displayed in Table \ref{Taba:DecayConstantVector}. The vectorial sector decouples from the other sectors and, in particular, does not contain any dependence on the tachyon field. Therefore, the vector meson decay constants are independent of the parameters $a_0$, $b_0$ and $\lambda$ and models A and B provide the same results. 
%%%%%%%%%%%%%%%%%%%%%
\begin{table}[ht]
\centering
\begin{tabular}{l |c|c|l}
\hline 
\hline
 & Models A and B & SW \cite{Karch:2006pv} &  Experimental \cite{Donoghue:1992dd} \\
\hline 
 $F_{\scriptscriptstyle{V_0}}^{\scriptscriptstyle{1/2}}$ & 276 &  261 & $346.2\pm 1.4$  \\
 $F_{\scriptscriptstyle{V_1}}^{\scriptscriptstyle{1/2}}$ & 341 &  & $433\pm 13$ \\
$F_{\scriptscriptstyle{V_2}}^{\scriptscriptstyle{1/2}}$ & 384 &  &  \\
\hline\hline
\end{tabular}
\caption{
The decay constants of the vector mesons (in MeV) obtained in our work (same result for models A and B), compared against the result obtained using the soft wall model (SW)~\cite{Karch:2006pv}. In order to compare with experimental results of \cite{Donoghue:1992dd}, we need to identify $F_{V}$ with $g_{\rho}$.
}
\label{Taba:DecayConstantVector}
\end{table}
%%%%%%%%%%%%%%%%%%%%%

\subsection{Scalar mesons}

We now calculate the decay constants of scalar mesons. The normalization condition for the wave functions, solutions of the eigenvalue problem, is given by (see \cite{Ballon-Bayona:2021ibm} for details on the derivation)
\noindent
\begin{equation}\label{Eq:NormCondScalar}
\int du\,\psi_{s_m}(u)\psi_{s_n}(u)=\int du\, e^{3A_s-\Phi}S_m(u)\,S_n(u)=\delta_{mn} \, ,
\end{equation}
\noindent 
where $S_n(u)=\,e^{-B_{S}}\psi_{s_{n}}(u)$. To get the normalization constant we first consider the asymptotic expansion of the normalizable solution, $S_n(u)$, close to the boundary,  $S_n(u)=N_{s_n}u^3+\cdots$, where $N_{s_n}$ is the normalization constant, which is obtained by plugging $S_{n}(u)$ in \eqref{Eq:NormCondScalar}. Then decay constants of the scalar mesons are given by  the following dictionary \cite{Ballon-Bayona:2021ibm}
\noindent
\begin{equation}
F_{s_n}=\zeta\,u\,e^{3A_s-\Phi}\partial_uS_n\bigg{|}_{u=\epsilon}=3\frac{\sqrt{N_c}}{2\pi}\,N_{s_n}.
\end{equation}
\noindent
%{\Blue [Comment] - Quem \'e $\zeta$? \'e $e^{-A}$? N\~ao seria melhor escrever em fun\c c\~ao de $A_s$?}

\begin{figure}[ht!]
\centering
\includegraphics[width=7cm]{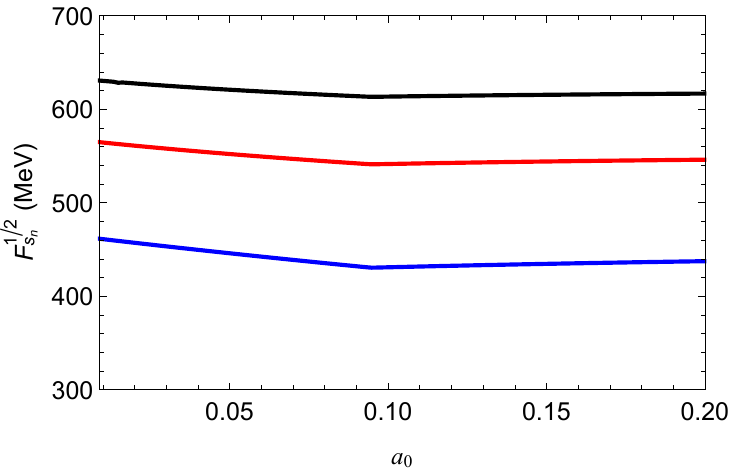}\hfill
\includegraphics[width=7cm]{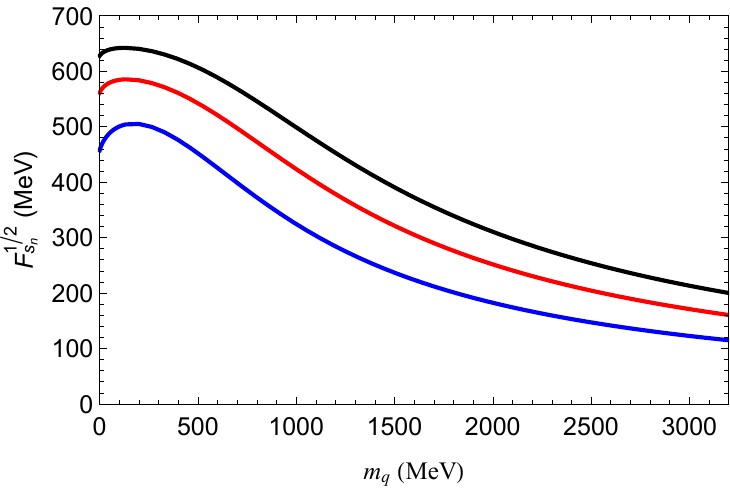}
\caption{
Left  panel: The decay constants of the scalar mesons as a function of $a_0$ in the chiral limit. Right panel: The decay constants of the scalar mesons as a function of the quark mass for $\lambda=27$, $b_0=1.7$ and $a_0=0.02$.
}
\label{Fig:DecayScalarMesons}
\end{figure}

 We first investigate the behavior of the scalar meson decay constants as a function of the parameter $a_0$ in the chiral limit with the other parameters fixed as in model A. Our numerical results are displayed on the left panel of Fig.~\ref{Fig:DecayScalarMesons}. As can be seen from the plot, the results show a smooth behavior of the decay constants in the region of interest, i.e., $a_0< a_0^c$, where $a_0^c\approx 0.0974$. However, the behavior of the decay constant   increases close to $a_0^c$ for the ground state while it decreases for the scalar resonances. 

In addition, we calculate the scalar meson decay constants as a function of the quark mass. Our numerical results are displayed on the right panel of Fig.~\ref{Fig:DecayScalarMesons}. As can be seen from the plot, the decay constant increases in the region of small quark mass until it reaches some maximum value  and then decreases when the quark mass grows. Finally, we calculate the decay constant using the final set of parameters for models A and B displayed in Table~\ref{tab:Parameters}. Our results for the scalar meson decay constants are displayed in Table \ref{Taba:DecayConstantS} compared against the results obtained in the soft wall model and a phenomenological model of QCD \cite{Gokalp:2001}.

%%%%%%%%%%%%%%%%%%%%%
\begin{table}[ht]
\centering
\begin{tabular}{l |c|c|c|c}
\hline 
\hline
 & Model A &Model B & SW \cite{Karch:2006pv}&  Other model \cite{Gokalp:2001}\\
\hline 
 $F_{\scriptscriptstyle{s_0}}^{\scriptscriptstyle{1/2}}$  & 463 & 459 &  420 & 425.3  \\
 $F_{\scriptscriptstyle{s_1}}^{\scriptscriptstyle{1/2}}$  & 565 & 562 & 499 &  \\
 $F_{\scriptscriptstyle{s_2}}^{\scriptscriptstyle{1/2}}$  & 630 & 628 & 552 &  \\
\hline\hline
\end{tabular}
\caption{
Decay constants of scalar mesons (in MeV) obtained in models A and B for the set of parameters given in Table \ref{tab:Parameters}, compared against the soft wall model ~\cite{Karch:2006pv} and the result obtained within a phenomenological model of QCD \cite{Gokalp:2001}.}
\label{Taba:DecayConstantS}
\end{table}
%%%%%%%%%%%%%%%%%%%%%

\subsection{Axial-vector mesons}

For the axial-vector mesons decay constants we follow the same procedure as before. Solving the eigenvalue problem \eqref{Eq:AVSchroEq}, one gets the wave functions, whose normalization condition is given by
\noindent
\begin{equation}
\int du\,\psi_{a_m}(u)\psi_{a_n}(u)=\int du\, e^{A_s-\Phi}a_m(z)\,a_n(u)=\delta_{mn}, \label{Eq:NormAxial}
\end{equation}
\noindent
where $a_n(u)=\,e^{-B_{A}}\psi_{a_{n}}(u)$ is the normalizable solution. Considering the expansion close to the boundary, $a_n(u)=N_{a_n} u^2 + \dots $, where $N_{a_n}$ is the normalization constant which is obtained plugging this result in  \eqref{Eq:NormAxial}. The axial-vector meson decay constants take the form \cite{Ballon-Bayona:2021ibm}
\noindent
\begin{equation}
F_{a_n}=\lim_{\epsilon\to 0}\frac{e^{A_s-\Phi}}{g_5}\,\partial_u a_n\bigg{|}_{z=\epsilon}=\frac{2}{g_5}N_{a_n}.
\end{equation}
\noindent

\begin{figure}[ht!]
\centering
\includegraphics[width=7cm]{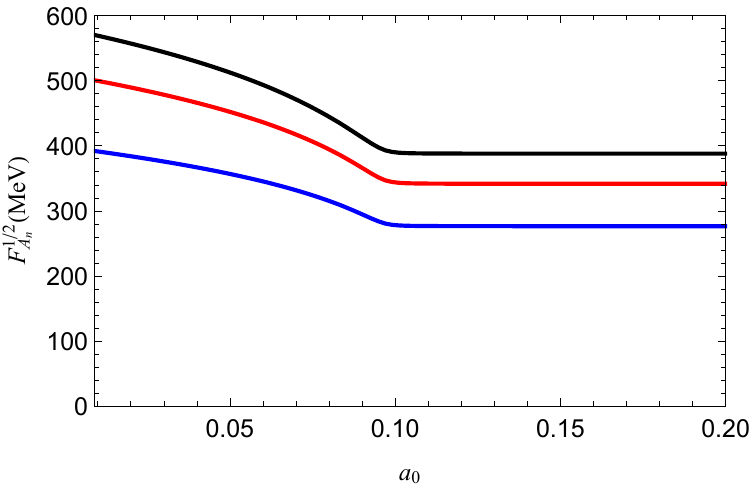}\hfill
\includegraphics[width=7cm]{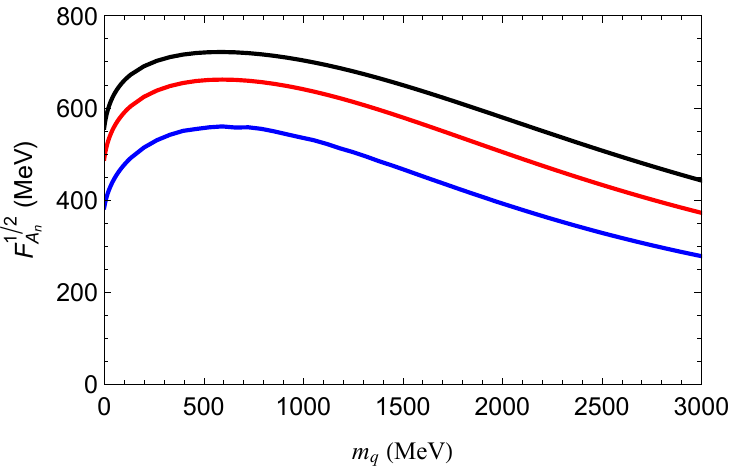}
\caption{
Left panel: The axial-vector decay constants as a function of $a_0$ in the chiral limit for $b_0=1.7$ and $\lambda=27$. Right panel: The axial-vector decay constants as a function of the quark mass for $\lambda=27$, $b_0=1.7$ and $a_0=0.02$.
}
\label{Fig:DecayAVMesons}
\end{figure}
As in the previous cases, the decay constants are obtained through the wave functions normalization constants. Firstly, we investigate the behavior of the decay constants as a function of $a_0$ in the chiral limit with the other parameters fixed as in model A. This is displayed on the left panel of Fig.~\ref{Fig:DecayAVMesons}. As can be seen from the plot, in the region of interest, i.e., $a_0< a_{0_c}$,  where $a_0^c\approx 0.0974$,  the decay constants decrease with the increasing of $a_0$ while they remain constant in the region corresponding to $a_0>a_{0_c}$. Fixing the value of $a_0$ and the other parameters as in Table \ref{tab:Parameters} for model A but allowing for the quark mass to vary, we investigate the behavior of the decay constants as a function of the quark mass. Our numerical results are displayed on the right panel of Fig. \ref{Fig:DecayAVMesons}. As can be seen from the plot, the axial-vector meson decay constants are finite in the chiral limit. Then, as the quark mass increases, they increase in the region of light quarks and decrease in the region of heavy-quarks. We finally calculate the decay constants for the set o parameters presented in Table \ref{tab:Parameters}. Our final results for the axial-vector meson decay constants are displayed in Table \ref{Taba:DecayConstantAV}, compared against the results obtained in the linear soft wall model \cite{Karch:2006pv} and the available experimental data \cite{Isgur:1988vm}.

%%%%%%%%%%%%%%%%%%%%%
\begin{table}[ht]
\centering
\begin{tabular}{l |c|c|c|l}
\hline 
\hline
 & Model A &Model B & SW \cite{Karch:2006pv} &  Extracted value \cite{Isgur:1988vm}  \\
\hline 
 $F_{\scriptscriptstyle{A_0}}^{\scriptscriptstyle{1/2}}$  & 395 & 581 & 261 & $433\pm 13$ \\
 $F_{\scriptscriptstyle{A_1}}^{\scriptscriptstyle{1/2}}$  & 503 & 763 &  &  \\
 $F_{\scriptscriptstyle{A_2}}^{\scriptscriptstyle{1/2}}$  & 572 & 889 &&  \\
\hline\hline
\end{tabular}
\caption{
Decay constants of axial-vector mesons (in MeV) obtained in models A and B, compared against the result obtained in the linear soft wall model ~\cite{Karch:2006pv} and the experimental data \cite{Isgur:1988vm}. In order to compare with experimental results of \cite{Isgur:1988vm}, we need to identify $F_{A^c}=f_{a_1}/\sqrt{2}$.
}
\label{Taba:DecayConstantAV}
\end{table}
%%%%%%%%%%%%%%%%%%%%%

\subsection{Pseudoscalar mesons}

To calculate the decay constants in the pseudoscalar sector we will use the procedure implemented in Ref.~\cite{Abidin:2009aj}, see also 
\cite{Ballon-Bayona:2014oma,Ballon-Bayona:2017bwk,Ballon-Bayona:2020qpq,Ballon-Bayona:2021ibm}. This procedure uses the normalization condition on the fields, which for the pseudoscalar mesons is given by
\noindent
\begin{equation}\label{Eq:NormPi}
\int_{\epsilon}^{\infty} du\,\frac{e^{A_s-\Phi}}{\beta(u)}\left(\partial_u\varphi_m\right)\left(\partial_u\varphi_n\right)=\frac{\delta_{mn}}{m_{\pi_n}^2},
\end{equation}
\noindent
where $\epsilon$ is a UV cutoff and $\varphi_n(u)$ is a solution of the coupled differential Eqs.~\eqref{varphieq} and \eqref{pieq}. It is worth mentioning that the normalization condition is different in comparison with the other sectors, meaning that, now, it depends on the mass of the states. The decay constants are calculated using the expression
\noindent
\begin{equation}\label{Eq:DecayPi1}
f_{\pi_n}=-\lim_{\epsilon\to 0}\frac{e^{A_s-\Phi}}{g_5}\partial_u\varphi_n(u)\bigg{|}_{u=\epsilon}.
\end{equation}
\noindent
One can write the last equation in terms of a normalization constant. Let us consider the expansion of the scalar field, $\varphi_n(u)$, close to the boundary, $\varphi_n=-N_{\pi_n}(u^2+\cdots)$, here $N_{\pi_n}$ is the normalization constant obtained from the normalization condition \eqref{Eq:NormPi}. Then, plugging this expression in \eqref{Eq:DecayPi1} and considering that the leading term is the AdS warp factor, $A_s=-\ln{u}$, we get
\noindent
\begin{equation}\label{Eq:DecayPi2}
f_{\pi_n}=\frac{2}{g_5}\,N_{\pi_n}.
\end{equation}
\noindent

\begin{figure}[ht!]
\centering
\includegraphics[width=7cm]{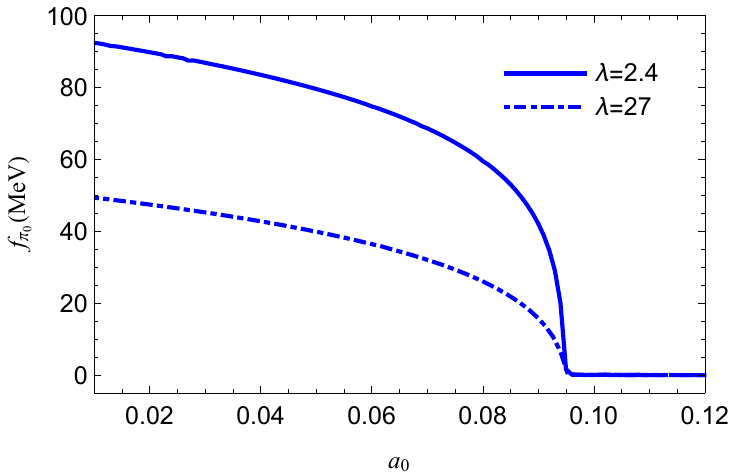}\hfill 
\includegraphics[width=7cm]{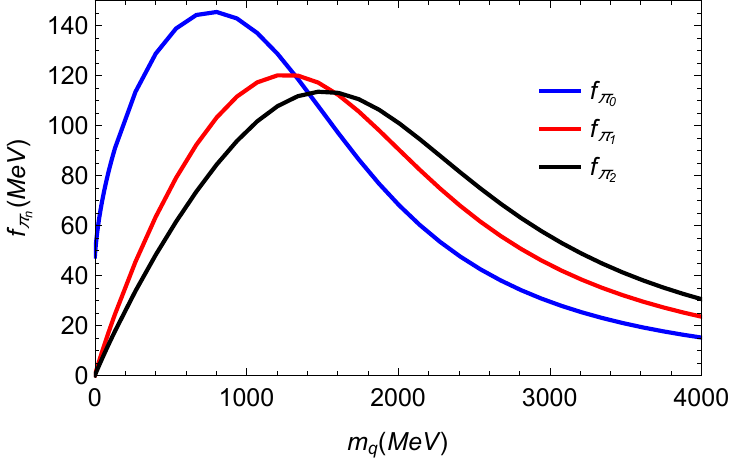}
\caption{
Left panel: Decay constant of the pseudoscalar ground state (the pion) in the chiral limit as a function of the parameter $a_0$, for $b_0=1.7$ and two values of $\lambda$, $\lambda =2.4$ (solid) and  $\lambda=27$ (dashed). Right panel: The decay constants of the pseudoscalar mesons as a function of the quark mass for $\lambda=27$, $b_0=1.7$ and $a_0=0.02$. 
}
\label{Fig:DecayPion}
\end{figure}

Once again, the decay constants are proportional to the normalization constants, the latter are obtained numerically integrating Eq.~\eqref{Eq:NormPi}. We first present our numerical results for the pseudoscalar ground state (the pion). The pion decay constant as a function of the parameter $a_0$ near the chiral limit, for different values of $\lambda$, is displayed on the left panel of Fig.~\ref{Fig:DecayPion}. As can be seen from the plot, the pion decay constant decreases with the increasing of $a_0$ in the region $a_0 < a_0^c$, where $a_0^c\approx 0.0974$, and becomes zero in the region $a_0 > a_0^c$. This is consistent with the fact that spontaneous chiral symmetry breaking occurs only in the region $a_0 < a_0^c$, as described in subsection \ref{Sec:XSB}.
Our results for the decay constants as a function of the quark mass are displayed on the right panel of Fig.~\ref{Fig:DecayPion}, where blue line represents the decay constant for the ground state (the pion), while red and black lines represent the decay constants for the first and second resonances. As can be seen from the plot, the decay constant for the ground state (the pion) remains finite in the chiral limit, while it vanishes for the excited states (pion resonances). Moreover, the hierarchy is inverted when we go from light to heavy-quarks. This very interesting behaviour for the pseudoscalar meson decay constant was previously obtained in the a non-linear extension of the soft wall model \cite{Ballon-Bayona:2021ibm}. 

Finally, we calculate the pseudoscalar meson decay constants for models A and B using the set of parameters given in Table \ref{tab:Parameters}. We remind the reader that the parameters $\lambda$ and $m_q$ A were fixed in model A using the masses of the first two pseudoscalar states while in model B they were fixed using the mass and pseudoscalar decay constant of the ground state (the pion). Our final results are displayed in Table \ref{Tab:DCPion}, where we compare against the available estimates from PDG \cite{Workman:2022ynf}.

%%%%%%%%%%%%%%%%%%%%%
\begin{table}[ht]
\centering
\begin{tabular}{l |c|c| l}
\hline 
\hline
 & Model A & Model B & PDG estimate \cite{Workman:2022ynf} \\
%  & & & & &   \\
\hline 
 $f_{\pi_0}$  & 51.5  & 92.1 &  $92.1\pm 0.8$  \\
 $f_{\pi_1}$  & 1.16  & 1.10 &   \\
 $f_{\pi_2}$  & 0.795 & 0.798 &   \\
\hline\hline
\end{tabular}
\caption{
The decay constants of pseudoscalar mesons (in MeV) obtained in models A and B, compared against the theoretical estimate made in PDG for  $f_{\pi^{+}}/\sqrt{2}$ \cite{Workman:2022ynf}.} 
\label{Tab:DCPion}
\end{table}
%%%%%%%%%%%%%%%%%%%%%

\subsection{Gell-Mann-Oakes-Renner relation}

An important test for a holographic QCD model for spontaneous chiral symmetry breaking is the Gell-Mann-Oakes-Renner (GOR) relation. We show in this subsection that our holographic model satisfies this important relation.  We will follow the analysis implemented in \cite{Abidin:2009aj, Ballon-Bayona:2021ibm}. The procedure starts with the normalization condition of the pseudoscalar mesons \eqref{Eq:NormPi}, then, we rewrite this equation in a form where the formula for the decay constant \eqref{Eq:DecayPi1} is evident. Moreover, note that in the UV the function $\Phi(u)$  is subleading compared against the warp factor $A_s\approx -\ln{u}$. Thus, considering $m=$n, one can write \eqref{Eq:NormPi} in the form
\noindent
\begin{equation}\label{eq:NormalCondGORv2}
\,m_{\pi_n}^2\int_{\epsilon}^{\infty} du\,\frac{e^{-3A_s-\Phi}}{v^2(u) } \left(\frac{e^{A_s}}{g_5}\partial_u\varphi_n\right)\left(\frac{e^{A_s}}{g_5}\partial_u\varphi_n\right)=1 \, ,
\end{equation}
where we replaced the definition of $\beta$. In the region near the chiral limit, i.e, $m_q \to 0$, it can be shown that the combination $e^{-3A_s-\Phi}/v^2(u)$ inside the integral becomes highly peaked near the boundary, see the discussion in \cite{Ballon-Bayona:2021ibm}. Meanwhile, terms like $e^{A_s}\partial_u\varphi_n$, inside the integral become constants, thus, they can be moved outside of the integral. However, this condition is satisfied only by the ground state because it has a finite decay constant in the chiral limit, see Fig.~\ref{Fig:DecayPion}. Then, for the ground state, the last equation can be written in the form (setting $m=n=0$)
\noindent
\begin{equation}\label{Eq:NorCondition2}
f_{\pi_0}^2m_{\pi_0}^2\int_{\epsilon}^{\infty} \frac{e^{-3A_s-\Phi}}{v^2(u)}du=1.
\end{equation}
\noindent
As discussed in \cite{Ballon-Bayona:2021ibm}, one can show that, near the chiral limit, the expression $e^{-3A_s-\Phi}/v^2(u)$ behaves like $e^{-3A_s-\Phi}/v^2(u)\sim u/(c_1+c_3u^2)^2$. Hence, after integrating, one gets
\noindent
\begin{equation}\label{Eq:NorCondition3}
\begin{split}
\int_{\epsilon}^{\infty} \frac{e^{-3A_s-\Phi}}{v^2(u)}du\approx\,&\frac{1}{2 c_1c_3}.
\end{split}
\end{equation}
\noindent
It is worth mentioning that the last result is only valid for small $c_1$. Plugging \eqref{Eq:NorCondition3} in \eqref{Eq:NorCondition2}, then, replacing  $c_1$ and $c_3$ by the quark mass $m_q$ and the chiral condensate $\Sigma$, using the relations $c_1=m_q\,\zeta$ and $c_3={\Sigma}/(2\zeta)$, we get the GOR relation
\noindent
\begin{equation}
f_{\pi_0}^2m_{\pi_0}^2=m_q\Sigma,
\end{equation}
\noindent
where the chiral condensate $\Sigma$ can be written as $\Sigma=\langle \bar{u}u\rangle+\langle \bar{d}d\rangle$, where $\langle \bar{u}u\rangle=\langle \bar{d}d\rangle=\sigma$ are the quark condensates for up and down quarks. Hence, the GOR relation can be written in the form
\noindent
\begin{equation}\label{Eq:GORrelation2}
f_{\pi_0}^2m_{\pi_0}^2=2m_q\sigma.
\end{equation}
\noindent

\begin{figure}[ht!]
\centering
\includegraphics[width=7cm]{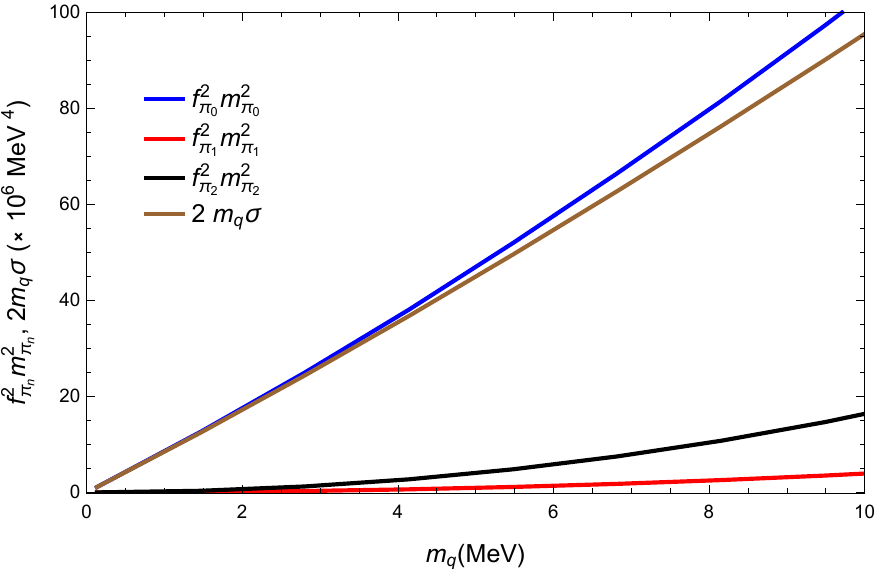}
\caption{
Our numerical results for the terms appearing on the left and right hand side of the GOR relation. This confirms numerically that the GOR relation is satisfied by the pseudoscalar ground state (the pion) in our holographic model.
}
\label{Fig:GOR}
\end{figure}

In order to verify the consistency of our derivation, we calculate the GOR relation numerically for model A. As we previously fixed the parameters, it is an relatively easy task to plot the left and right hand side terms appearing in the GOR relation as a function of the quark mass. Our numerical results are displayed in Fig.~\ref{Fig:GOR}, where solid blue line represents the results for $f_{\pi_0}^2m_{\pi_0}^2$, while solid brown line represents the results for $2m_q\sigma$. As can be seen from the plot, the curves overlap near the chiral limit. We also show in the figure that the product $f_{\pi_n}^2m_{\pi_n}^2$ for $n=1$ and $n=2$ (the first two excited states) do not match with the function  $2m_q\sigma$ and therefore pion resonances do not satisfy the GOR relation, as expected in QCD. 

We finish this work presenting our estimate for the chiral condensate using the final set of parameters displayed in Table \ref{tab:Parameters}. Using the relation $\sigma=c_3\Lambda^{3}\zeta$ and evaluating numerically the UV coefficient $c_3$ we find  $\sigma=(160\,\text{MeV})^3$ for model A and $\sigma=(240\,\text{MeV})^3$ for model B.  These results are of the same order as those obtained in lattice QCD, see for instance \cite{Shuryak:1988ck, Giusti:1998wy, Fukaya:2009fh,McNeile:2012xh}.

\section{Conclusions}
\label{Sec:Conclusions}

In this paper, we built a holographic QCD model that describes spontaneous chiral symmetry breaking and the meson spectrum in a confined background. The confined background was described by a solution of the Einstein-dilaton equations leading to a warp factor in the string frame that satisfies the confinement criterion of  Ref.~\cite{Kinar:1998vq}. The 4d currents and the quark mass operator associated with chiral symmetry breaking and the meson spectrum were described by a 5d probe action embedded on the gravitational background. We introduced the function $f(\Phi)=1/(1+b_0\Phi+a_0\Phi^2) $ for the dilaton coupling to the tachyon potential in order to get a solution for the tachyon that is regular in the IR and compatible with the AdS/CFT dictionary for the quark mass operator. From the tachyon solution close to the boundary we extracted the source and VEV parameters  which are related to the quark mass and the chiral condensate. One can think about the function $f(\Phi)$ as a regulator for the tachyon potential, which close to the boundary (UV) allows for the formation of a chiral condensate and far from the boundary (IR) guarantees that confinement is the dominant effect. The last property is important to guarantee asymptotically linear Regge trajectories with a mass scale related to the quadratic dilaton behavior in Einstein-dilaton gravity. This is a generalization of the phenomenological soft wall models that allows for a connection between confinement and hadron mass generation.

The model allowed us to describe spontaneous chiral symmetry breaking in the region of $a_0\leq a_{0_c}$, where we find a lightest state in the pseudoscalar sector, which we identify as the pion. We fixed the parameter $b_0$ to guarantee a positive chiral condensate for any value of the quark mass. Then, we calculated the spectrum for vector mesons and used the experimental result for the ground state, the $\rho$ meson, to fix the parameter $\Lambda$. Next, we focused on the scalar mesons. First, we analyzed the  instabilities arising in the spectrum when considering a trivial profile for the tachyon field. We also showed how those instabilities are eliminated when one moves to the non-trivial profile for the tachyon field, obtained solving the non-linear differential equation for the tachyon field. We fixed the parameter $a_0$ using the experimental result for the mass of the first excited state in the scalar sector, the $f_0(980)$ meson. The remaining parameters $\lambda$ (coupling of the tachyon potential) and $m_q$ (quark mass) were fixed in two different ways leading to models A and B. In model A we fixed the masses of first two states in the pseudoscalar sector (the pion and the first pion resonance) while in model B we fixed the mass and decay constant of the ground state (the pion).   
Finally, we observed that in the chiral limit the masses of the scalar mesons and axial-vector mesons remain finite. For the pseudoscalar mesons we found that mass of the ground state becomes zero in the chiral limit, while the mass of the resonances remain finite. This is consistent with the spontaneous breaking of chiral symmetry in QCD. 
We also observed an increasing of the meson masses with the increasing of the quark mass for the scalar, axial-vector and pseudoscalar mesons. Finally, for the set of parameters given in Table \ref{tab:Parameters}  we calculated the meson spectrum of all sectors (vectorial, axial-vectorial, scalar and pseudoscalar) and compared against the available  experimental results.

We also calculated the decay constants, which are related to the normalization condition of the wave functions. In our setup the vector meson decay constants do not depend on the quark mass. In turn, the decay constants for the scalar, axial-vector and pseudoscalar mesons do depend on the quark mass.   We also observed that the decay constants in the scalar sector are finite in the chiral limit, then, they increase as the quark mass increases reaching maximum values to finally decrease with the increasing of the quark mass, as shown in the right panel of Fig.~\ref{Fig:DecayScalarMesons}.. It is worth mentioning that the hierarchy between the masses in the scalar sector does not change. The axial-vector meson decay constants also depend on the quark mass and remain finite in the chiral limit, increasing with the quark mass in the region of light quarks, then, decreasing in the region of heavy quarks, as shown in the right panel of Fig.~\ref{Fig:DecayAVMesons}. The hierarchy between the masses of axial-vector mesons also does not change. In the pseudoscalar sector the decay constant of the ground state, i.e., the pion, remains finite in the chiral limit, while the decay constants for the excited states become zero. We also observed that the decay constants increase up to a maximum value, then, the hierarchy changes, to finally decrease with the increasing of the quark mass, as shown on the right panel of Fig.~\ref{Fig:DecayPion}. Finally, we obtained the Gell-Mann-Oakes-Renner relation for the pseudoscalar ground state (the pion) using our numerical results for the mass, decay constant and the chiral condensate.

In conclusion, we have built a holographic model that describes confinement and the spontaneous breaking of chiral symmetry and also provides a good description of the meson spectrum and decay constants.  Of course, this is a first step to build up a more complete holographic model and refinements will be added in the future. In the next stage of this project we shall address the calculation of the strong couplings between the mesons, as implemented in the context of the holographic hard wall model in \cite{Ballon-Bayona:2014oma}, the melting of the mesons, in particular the pion, following the ideas of \cite{ Miranda:2009uw,Mamani:2013ssa,Bartz:2016ufc,Mamani:2018uxf,Mamani:2022qnf}.

\section*{Acknowledgments}

The authors would like to acknowledge Carlisson Miller for stimulating discussions in the early stages of this project, we also thank Diego Rodrigues for discussions along the development of this work. The work of A.B-B is partially funded by Conselho Nacional de Desenvolvimento Cient\'\i fico e Tecnol\'ogico (CNPq, Brazil), grant No. 314000/2021-6, and Coordena\c{c}\~ao de Aperfei\c{c}oamento do Pessoal de N\'ivel Superior (CAPES, Brazil), Finance Code 001.
This work is a part of the project INCT-FNA \#464898/2014-5 (W.P. and T.F.).
The work of T.F. is partially financed  by CNPq-Brazil, under the grant No. 308486/2015-3, and by Funda\c c\~ao de Amparo \`a Pesquisa do Estado de S\~ao Paulo (FAPESP) under the Thematic  grants    No. 2017/05660-0  and No. 2019/07767-1.
W. d. P. acknowledges the partial support of CNPQ under Grant No. 313030/2021-9 and the partial support of CAPES under Grant No. 88881.309870/2018-01.

\appendix

\section{Numerical analysis}
\label{Sec:AddNumerical}

\begin{figure}[ht!]
\centering
\includegraphics[width=7cm]{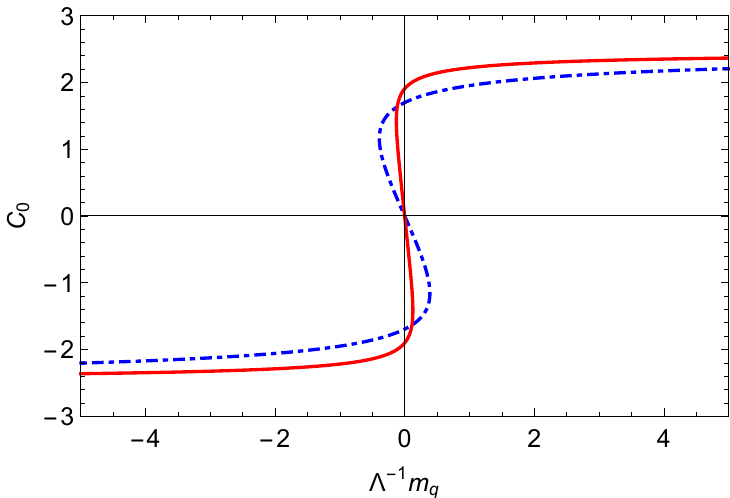}\hfill
\includegraphics[width=7cm]{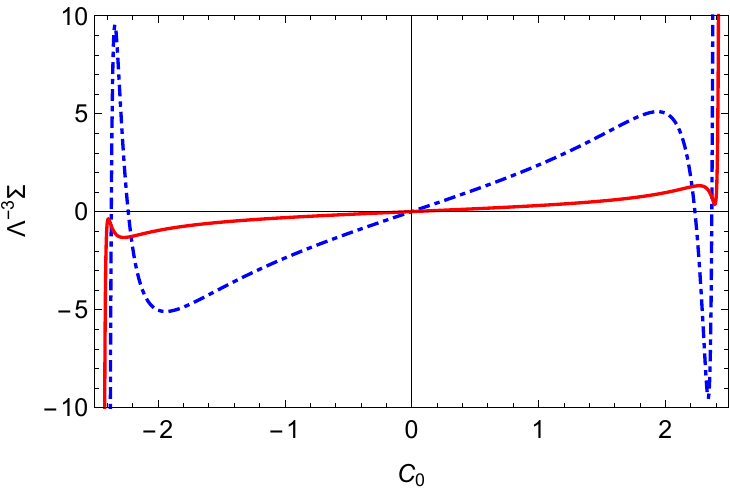}
\caption{
Left panel: $C_0$ as a function of the dimensionless quark mass. Right panel: The dimensionless chiral condensate as a function of $C_0$. These results were obtained for $a_0=1$ and $b_0=0$ (dotted blue line), and $a_0=0.02$ and $b_0=1.7$ (solid red line), considering $\lambda=1$.
}
\label{Fig:C1C0C0C3MII}
\end{figure}

Here we write additional details concerning the numerical analysis we implemented to calculate the chiral condensate as a function of the quark mass, as well as the dependence between the parameter $C_0$ and quark mass. The chiral condensate as a function of the quark mass can be seen in Fig.~\ref{Fig:C1C0C0C3MII}, where dotted blue line was obtained for $a_0=1$ and $b_0=0$, while red solid line was obtained for $a_0=0.02$ and $b_0=1.7$. Note that we have a non-zero value for the chiral condensate in the chiral limit. From the figure, one can conclude that there is a symmetry between the first and third quadrant of the plane, i.e., $(m_q, \Sigma)\to (-m_q, -\Sigma)$ and between the second and fourth quadrant of the plane, i.e., $(-m_q, \Sigma)\to (m_q, -\Sigma)$.

\begin{figure}[ht!]
\centering
\includegraphics[width=7cm]{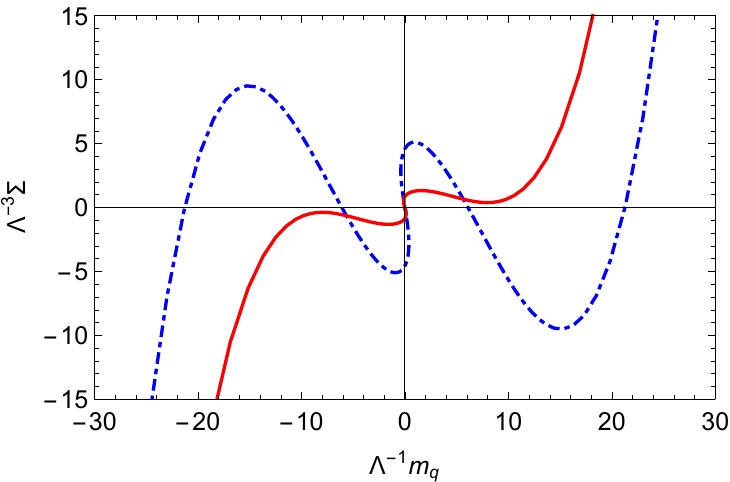}
\caption{
The figure shows the numerical results for the the dimensionless chiral condensate as a function of the dimensionless quarks mass for $a_0=1$ and $b_0=0$ (dotted blue line), and $a_0=0.02$ and $b_0=1.7$ (solid red line), considering $\lambda=1$.
}
\label{Fig:C1C3MII}
\end{figure}

We also calculate the dependence of the parameter $C_0$, as a function of the dimensionless quark mass in the left panel of Fig.~\ref{Fig:C1C0C0C3MII}. As can be seen, it is finite in the chiral limit. Then, it increases slowly in the physical region, i.e., $m_q>0$. This is true for the set of parameters considered in Fig.~\ref{Fig:C1C3MII}. We also calculated the chiral condensate as a function of the parameter $C_0$, see right panel in Fig.~\ref{Fig:C1C0C0C3MII}. As can be seen, for the set of parameters $a_0=1$ and $b_0=0$, the chiral condensate has five zeros, while for $a_0=1$ and $b_0=0$, it has one zero.

\section{Alternative models}

Here we write other models we have investigated along the development of this project. The models provide results qualitatively equivalents to those presented in this paper. For the warp factor in Einstein frame, we consider:
\noindent
\begin{equation}
\begin{split}
A_{II}=\,&-\ln{u}-\frac{u^4}{1+\xi^2 u^2},\qquad\qquad\qquad\qquad (\text{Model II})\\
A_{III}=\,&-\ln{u}-u^2\qquad\qquad\qquad\qquad\qquad (\text{Model III})
\end{split}
\end{equation}
\noindent

The main characteristic of model III is that one gets an analytic function for the dilaton field. Meanwhile, in model II the dilaton becomes numeric. The advantage of model III is that it spend less time than model II to get numerical results within a controlled region of numerical uncertainty.

\section{Divergent tachyon: Numerical solution}
\label{Sec:DiverModel}

In this section we consider the simplest possible solution within the dilaton-gravity solution and the tachyon differential equation which supports a divergent solution. The simplest solution providing confinement and linear Regge behavior is given by the analytic warp factor
\noindent
\begin{equation}\label{Eq:GKNWarp}
A=-\ln{\left(u\right)}-u^2.
\end{equation}
\noindent
By plugging this function in Eq.~\eqref{Eq:EinsteinDil} it is possible to get an analytic solution for the dilaton field
\noindent
\begin{equation}\label{Eq:GKNDil}
\Phi=\frac{3 u}{4}\sqrt{6+4u^2}+\frac{9}{4}\ln{\left(\frac{2u+\sqrt{6+4u^2}}{\sqrt{6}}\right)}
\end{equation}
\noindent

The analytic warp factor \eqref{Eq:GKNWarp} model was proposed in Appendix G of ref.~\cite{Gursoy:2007er}. Once we have the gravitational background we now turn our attention to the tachyon field. Considering $f(\Phi)= 1$ and $\lambda=0$ Eq.~\eqref{Eq:TachyonEq} reduces to \noindent
\begin{equation}\label{Eq:DivTachyonEq}
v''(u)+\left(3A_s'(u)-\Phi'(u)\right)v'(u)+3e^{2A_s}v(u)=0.
\end{equation}

The asymptotic solution close to the boundary is given by
\noindent
\begin{equation}\label{Eq:UVSolDivTachyon}
\begin{split}
&v=c_1\,u+15\sqrt{\frac{3}{2}}c_1\,u^2+c_3\,u^3-\frac{429}{2}c_1\,u^3\ln{u^3}+\cdots.\qquad u\to 0
\end{split}
\end{equation}
\noindent

\begin{figure}[ht!]
\centering
\includegraphics[width=7cm]{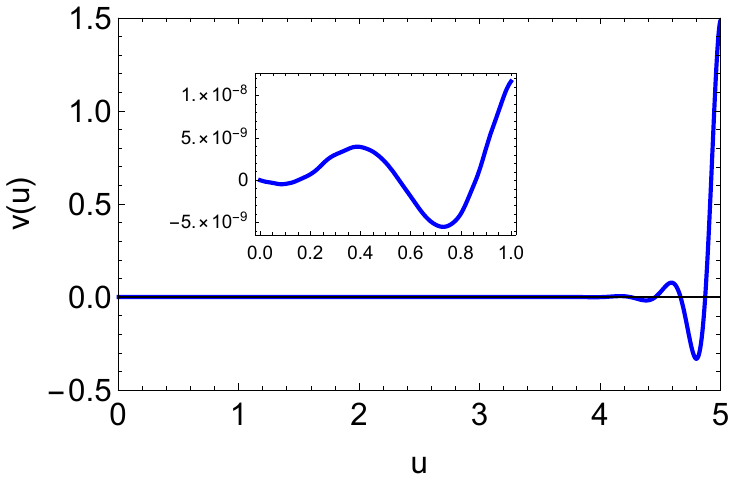}\hfill 
\includegraphics[width=7cm]{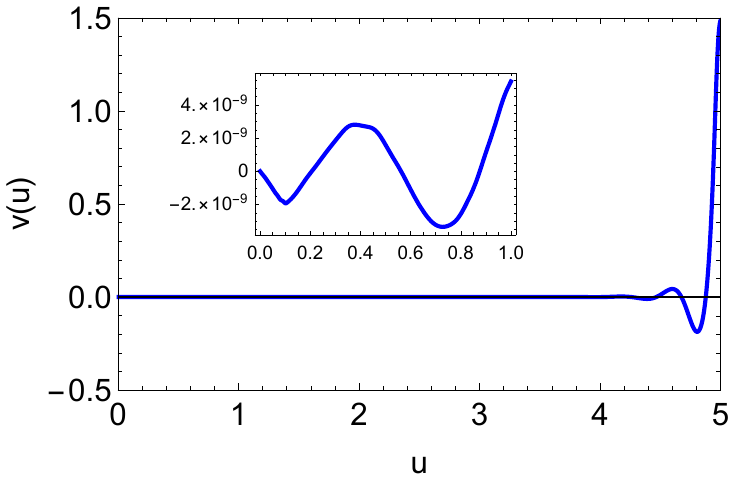}
\caption{
Left panel: The profile of the tachyon obtained solving numerically Eq.~\eqref{Eq:DivTachyonEq}. Right panel: The profile of the tachyon obtained solving numerically Eq.~\eqref{Eq:DivTachyonEq2}.
}
\label{Fig:DivTachyon}
\end{figure}

In turn, the leading term of the asymptotic solution in the IR is given by
\noindent
\begin{equation}\label{Eq:IRSolDivTachyon}
v=C_0\,e^{u}. \qquad u \to \infty
\end{equation}
\noindent
We now solve this equation numerically. As the solution in the IR depends on one parameter we solve the differential equation using as ``initial condition'' the asymptotic solution in the IR. Our numerical results for the profile of the tachyon are displayed in the left panel of Fig.~\ref{Fig:DivTachyon}. As can be seen, the solution has an oscillatory behavior. This behavior does not allow to read off the coefficients $c_1$ and $c_3$ of the asymptotic solution \eqref{Eq:UVSolDivTachyon}.

On the other hand, we can repeat the same analysis using the equation derived in Section 6.7.1 of Ref.~\cite{Gursoy:2007er}, see Eq.~(6.63). Thus, the differential equation for the tachyon is 
\noindent
\begin{equation}\label{Eq:DivTachyonEq2}
v''(u)+(3 A_s'-\Phi')v'(u)+3\,e^{2A_s}v(u)+e^{-2A_s}\left(4 A_s'-\Phi'\right)(v'(u))^3+3v(u)\,(v'(u))^2=0.
\end{equation}
\noindent
As we have the gravitational background \eqref{Eq:GKNWarp} and \eqref{Eq:GKNDil} we may solve the last equation numerically. The asymptotic solution in the IR is given by the divergent \eqref{Eq:IRSolDivTachyon}. Our numerical results for the tachyon field are displayed in the right panel of Fig.~\ref{Fig:DivTachyon}.

%\bibliography{HologXSB_InterpolatingDilaton}
%\bibliographystyle{abbrv}%{apalike}%{h-physrev}
%\bibliographystyle{ieeetr}
%\bibliographystyle{plain}
%\bibliographystyle{unsrt}
%\bibliographystyle{apsrev4-2}
%\bibliographystyle{apalike2}

\bibliographystyle{utphys}
\bibliography{XSB_Last}

\providecommand{\href}[2]{#2}\begingroup\raggedright\begin{thebibliography}{10}

\bibitem{RevModPhys.81.1015}
Y.~Nambu, ``Nobel lecture: Spontaneous symmetry breaking in particle physics: A
  case of cross fertilization,''
  \href{http://dx.doi.org/10.1103/RevModPhys.81.1015}{{\em Rev. Mod. Phys.}
  {\bfseries 81} (Jul, 2009) 1015--1018}.
  \url{https://link.aps.org/doi/10.1103/RevModPhys.81.1015}.

\bibitem{PhysRev.127.965}
J.~Goldstone, A.~Salam, and S.~Weinberg, ``Broken symmetries,''
  \href{http://dx.doi.org/10.1103/PhysRev.127.965}{{\em Phys. Rev.} {\bfseries
  127} (Aug, 1962) 965--970}.
  \url{https://link.aps.org/doi/10.1103/PhysRev.127.965}.

\bibitem{PhysRev.128.2462}
M.~Baker and S.~L. Glashow, ``Spontaneous breakdown of elementary particle
  symmetries,'' \href{http://dx.doi.org/10.1103/PhysRev.128.2462}{{\em Phys.
  Rev.} {\bfseries 128} (Dec, 1962) 2462--2471}.
  \url{https://link.aps.org/doi/10.1103/PhysRev.128.2462}.

\bibitem{Tong2018}
D.~Tong, ``{Gauge theory},''  (2018) .
  \url{http://www.damtp.cam.ac.uk/user/tong/gaugetheory/gt.pdf}.

\bibitem{Coleman:1980mx}
S.~R. Coleman and E.~Witten, ``{Chiral Symmetry Breakdown in Large N
  Chromodynamics},'' \href{http://dx.doi.org/10.1103/PhysRevLett.45.100}{{\em
  Phys. Rev. Lett.} {\bfseries 45} (1980) 100}.

\bibitem{Maldacena:1997re}
J.~M. Maldacena, ``{The Large N limit of superconformal field theories and
  supergravity},'' \href{http://dx.doi.org/10.1023/A:1026654312961}{{\em Adv.
  Theor. Math. Phys.} {\bfseries 2} (1998) 231--252},
  \href{http://arxiv.org/abs/hep-th/9711200}{{\ttfamily arXiv:hep-th/9711200}}.

\bibitem{Maldacena:2000yy}
J.~M. Maldacena and C.~Nunez, ``{Towards the large N limit of pure N=1
  superYang-Mills},'' \href{http://dx.doi.org/10.1103/PhysRevLett.86.588}{{\em
  Phys. Rev. Lett.} {\bfseries 86} (2001) 588--591},
  \href{http://arxiv.org/abs/hep-th/0008001}{{\ttfamily arXiv:hep-th/0008001}}.

\bibitem{Gubser:1998bc}
S.~S. Gubser, I.~R. Klebanov, and A.~M. Polyakov, ``{Gauge theory correlators
  from noncritical string theory},''
  \href{http://dx.doi.org/10.1016/S0370-2693(98)00377-3}{{\em Phys. Lett. B}
  {\bfseries 428} (1998) 105--114},
  \href{http://arxiv.org/abs/hep-th/9802109}{{\ttfamily arXiv:hep-th/9802109}}.

\bibitem{Witten:1998qj}
E.~Witten, ``{Anti-de Sitter space and holography},''
  \href{http://dx.doi.org/10.4310/ATMP.1998.v2.n2.a2}{{\em Adv. Theor. Math.
  Phys.} {\bfseries 2} (1998) 253--291},
  \href{http://arxiv.org/abs/hep-th/9802150}{{\ttfamily arXiv:hep-th/9802150}}.

\bibitem{BoschiFilho:2002ta}
H.~Boschi-Filho and N.~R.~F. Braga, ``{QCD / string holographic mapping and
  glueball mass spectrum},''
  \href{http://dx.doi.org/10.1140/epjc/s2003-01526-4}{{\em Eur. Phys. J. C}
  {\bfseries 32} (2004) 529--533},
  \href{http://arxiv.org/abs/hep-th/0209080}{{\ttfamily arXiv:hep-th/0209080}}.

\bibitem{BoschiFilho:2002vd}
H.~Boschi-Filho and N.~R.~F. Braga, ``{Gauge / string duality and scalar
  glueball mass ratios},''
  \href{http://dx.doi.org/10.1088/1126-6708/2003/05/009}{{\em JHEP} {\bfseries
  05} (2003) 009}, \href{http://arxiv.org/abs/hep-th/0212207}{{\ttfamily
  arXiv:hep-th/0212207}}.

\bibitem{Erlich:2005qh}
J.~Erlich, E.~Katz, D.~T. Son, and M.~A. Stephanov, ``{QCD and a holographic
  model of hadrons},''
  \href{http://dx.doi.org/10.1103/PhysRevLett.95.261602}{{\em Phys. Rev. Lett.}
  {\bfseries 95} (2005) 261602},
  \href{http://arxiv.org/abs/hep-ph/0501128}{{\ttfamily arXiv:hep-ph/0501128}}.

\bibitem{Karch:2006pv}
A.~Karch, E.~Katz, D.~T. Son, and M.~A. Stephanov, ``{Linear confinement and
  AdS/QCD},'' \href{http://dx.doi.org/10.1103/PhysRevD.74.015005}{{\em Phys.
  Rev. D} {\bfseries 74} (2006) 015005},
  \href{http://arxiv.org/abs/hep-ph/0602229}{{\ttfamily arXiv:hep-ph/0602229}}.

\bibitem{Karch:2002sh}
A.~Karch and E.~Katz, ``{Adding flavor to AdS / CFT},''
  \href{http://dx.doi.org/10.1088/1126-6708/2002/06/043}{{\em JHEP} {\bfseries
  06} (2002) 043}, \href{http://arxiv.org/abs/hep-th/0205236}{{\ttfamily
  arXiv:hep-th/0205236}}.

\bibitem{Evans:2004ia}
N.~J. Evans and J.~P. Shock, ``{Chiral dynamics from AdS space},''
  \href{http://dx.doi.org/10.1103/PhysRevD.70.046002}{{\em Phys. Rev. D}
  {\bfseries 70} (2004) 046002},
  \href{http://arxiv.org/abs/hep-th/0403279}{{\ttfamily arXiv:hep-th/0403279}}.

\bibitem{Sakai:2004cn}
T.~Sakai and S.~Sugimoto, ``{Low energy hadron physics in holographic QCD},''
  \href{http://dx.doi.org/10.1143/PTP.113.843}{{\em Prog. Theor. Phys.}
  {\bfseries 113} (2005) 843--882},
  \href{http://arxiv.org/abs/hep-th/0412141}{{\ttfamily arXiv:hep-th/0412141}}.

\bibitem{deTeramond:2005su}
G.~F. de~Teramond and S.~J. Brodsky, ``{Hadronic spectrum of a holographic dual
  of QCD},'' \href{http://dx.doi.org/10.1103/PhysRevLett.94.201601}{{\em Phys.
  Rev. Lett.} {\bfseries 94} (2005) 201601},
  \href{http://arxiv.org/abs/hep-th/0501022}{{\ttfamily arXiv:hep-th/0501022}}.

\bibitem{Sakai:2005yt}
T.~Sakai and S.~Sugimoto, ``{More on a holographic dual of QCD},''
  \href{http://dx.doi.org/10.1143/PTP.114.1083}{{\em Prog. Theor. Phys.}
  {\bfseries 114} (2005) 1083--1118},
  \href{http://arxiv.org/abs/hep-th/0507073}{{\ttfamily arXiv:hep-th/0507073}}.

\bibitem{Andreev:2006ct}
O.~Andreev and V.~I. Zakharov, ``{Heavy-quark potentials and AdS/QCD},''
  \href{http://dx.doi.org/10.1103/PhysRevD.74.025023}{{\em Phys. Rev. D}
  {\bfseries 74} (2006) 025023},
  \href{http://arxiv.org/abs/hep-ph/0604204}{{\ttfamily arXiv:hep-ph/0604204}}.

\bibitem{Grigoryan:2007vg}
H.~R. Grigoryan and A.~V. Radyushkin, ``{Form Factors and Wave Functions of
  Vector Mesons in Holographic QCD},''
  \href{http://dx.doi.org/10.1016/j.physletb.2007.05.044}{{\em Phys. Lett. B}
  {\bfseries 650} (2007) 421--427},
  \href{http://arxiv.org/abs/hep-ph/0703069}{{\ttfamily arXiv:hep-ph/0703069}}.

\bibitem{Gursoy:2007er}
U.~Gursoy, E.~Kiritsis, and F.~Nitti, ``{Exploring improved holographic
  theories for QCD: Part II},''
  \href{http://dx.doi.org/10.1088/1126-6708/2008/02/019}{{\em JHEP} {\bfseries
  02} (2008) 019}, \href{http://arxiv.org/abs/0707.1349}{{\ttfamily
  arXiv:0707.1349 [hep-th]}}.

\bibitem{Cherman:2008eh}
A.~Cherman, T.~D. Cohen, and E.~S. Werbos, ``{The Chiral condensate in
  holographic models of QCD},''
  \href{http://dx.doi.org/10.1103/PhysRevC.79.045203}{{\em Phys. Rev. C}
  {\bfseries 79} (2009) 045203},
  \href{http://arxiv.org/abs/0804.1096}{{\ttfamily arXiv:0804.1096 [hep-ph]}}.

\bibitem{dePaula:2008fp}
W.~de~Paula, T.~Frederico, H.~Forkel, and M.~Beyer, ``{Dynamical AdS/QCD with
  area-law confinement and linear Regge trajectories},''
  \href{http://dx.doi.org/10.1103/PhysRevD.79.075019}{{\em Phys. Rev. D}
  {\bfseries 79} (2009) 075019},
  \href{http://arxiv.org/abs/0806.3830}{{\ttfamily arXiv:0806.3830 [hep-ph]}}.

\bibitem{Colangelo:2008us}
P.~Colangelo, F.~De~Fazio, F.~Giannuzzi, F.~Jugeau, and S.~Nicotri, ``{Light
  scalar mesons in the soft-wall model of AdS/QCD},''
  \href{http://dx.doi.org/10.1103/PhysRevD.78.055009}{{\em Phys. Rev. D}
  {\bfseries 78} (2008) 055009},
  \href{http://arxiv.org/abs/0807.1054}{{\ttfamily arXiv:0807.1054 [hep-ph]}}.

\bibitem{Vega:2008af}
A.~Vega and I.~Schmidt, ``{Scalar hadrons in AdS(5) x S**5},''
  \href{http://dx.doi.org/10.1103/PhysRevD.78.017703}{{\em Phys. Rev. D}
  {\bfseries 78} (2008) 017703},
  \href{http://arxiv.org/abs/0806.2267}{{\ttfamily arXiv:0806.2267 [hep-ph]}}.

\bibitem{Abidin:2009hr}
Z.~Abidin and C.~E. Carlson, ``{Nucleon electromagnetic and gravitational form
  factors from holography},''
  \href{http://dx.doi.org/10.1103/PhysRevD.79.115003}{{\em Phys. Rev. D}
  {\bfseries 79} (2009) 115003},
  \href{http://arxiv.org/abs/0903.4818}{{\ttfamily arXiv:0903.4818 [hep-ph]}}.

\bibitem{Abidin:2009aj}
Z.~Abidin and C.~E. Carlson, ``{Strange hadrons and kaon-to-pion transition
  form factors from holography},''
  \href{http://dx.doi.org/10.1103/PhysRevD.80.115010}{{\em Phys. Rev. D}
  {\bfseries 80} (2009) 115010},
  \href{http://arxiv.org/abs/0908.2452}{{\ttfamily arXiv:0908.2452 [hep-ph]}}.

\bibitem{dePaula:2009za}
W.~de~Paula and T.~Frederico, ``{Scalar mesons within a dynamical holographic
  QCD model},'' \href{http://dx.doi.org/10.1016/j.physletb.2010.08.045}{{\em
  Phys. Lett. B} {\bfseries 693} (2010) 287--291},
  \href{http://arxiv.org/abs/0908.4282}{{\ttfamily arXiv:0908.4282 [hep-ph]}}.

\bibitem{Bianchi:2010cy}
M.~Bianchi and W.~de~Paula, ``{On Exact Symmetries and Massless Vectors in
  Holographic Flows and other Flux Vacua},''
  \href{http://dx.doi.org/10.1007/JHEP04(2010)113}{{\em JHEP} {\bfseries 04}
  (2010) 113}, \href{http://arxiv.org/abs/1003.2536}{{\ttfamily arXiv:1003.2536
  [hep-th]}}.

\bibitem{Gherghetta:2009ac}
T.~Gherghetta, J.~I. Kapusta, and T.~M. Kelley, ``{Chiral symmetry breaking in
  the soft-wall AdS/QCD model},''
  \href{http://dx.doi.org/10.1103/PhysRevD.79.076003}{{\em Phys. Rev. D}
  {\bfseries 79} (2009) 076003},
  \href{http://arxiv.org/abs/0902.1998}{{\ttfamily arXiv:0902.1998 [hep-ph]}}.

\bibitem{Jarvinen:2011qe}
M.~Jarvinen and E.~Kiritsis, ``{Holographic Models for QCD in the Veneziano
  Limit},'' \href{http://dx.doi.org/10.1007/JHEP03(2012)002}{{\em JHEP}
  {\bfseries 03} (2012) 002}, \href{http://arxiv.org/abs/1112.1261}{{\ttfamily
  arXiv:1112.1261 [hep-ph]}}.

\bibitem{Li:2013oda}
D.~Li and M.~Huang, ``{Dynamical holographic QCD model for glueball and light
  meson spectra},'' \href{http://dx.doi.org/10.1007/JHEP11(2013)088}{{\em JHEP}
  {\bfseries 11} (2013) 088}, \href{http://arxiv.org/abs/1303.6929}{{\ttfamily
  arXiv:1303.6929 [hep-ph]}}.

\bibitem{Arean:2013tja}
D.~Are\'an, I.~Iatrakis, M.~J\"arvinen, and E.~Kiritsis, ``{The discontinuities
  of conformal transitions and mass spectra of V-QCD},''
  \href{http://dx.doi.org/10.1007/JHEP11(2013)068}{{\em JHEP} {\bfseries 11}
  (2013) 068}, \href{http://arxiv.org/abs/1309.2286}{{\ttfamily arXiv:1309.2286
  [hep-ph]}}.

\bibitem{Alho:2013dka}
T.~Alho, N.~Evans, and K.~Tuominen, ``{Dynamic AdS/QCD and the Spectrum of
  Walking Gauge Theories},''
  \href{http://dx.doi.org/10.1103/PhysRevD.88.105016}{{\em Phys. Rev. D}
  {\bfseries 88} (2013) 105016},
  \href{http://arxiv.org/abs/1307.4896}{{\ttfamily arXiv:1307.4896 [hep-ph]}}.

\bibitem{Ballon-Bayona:2014oma}
A.~Ballon-Bayona, G.~a. Krein, and C.~Miller, ``{Decay constants of the pion
  and its excitations in holographic QCD},''
  \href{http://dx.doi.org/10.1103/PhysRevD.91.065024}{{\em Phys. Rev. D}
  {\bfseries 91} (2015) 065024},
  \href{http://arxiv.org/abs/1412.7505}{{\ttfamily arXiv:1412.7505 [hep-ph]}}.

\bibitem{Chelabi:2015gpc}
K.~Chelabi, Z.~Fang, M.~Huang, D.~Li, and Y.-L. Wu, ``{Chiral Phase Transition
  in the Soft-Wall Model of AdS/QCD},''
  \href{http://dx.doi.org/10.1007/JHEP04(2016)036}{{\em JHEP} {\bfseries 04}
  (2016) 036}, \href{http://arxiv.org/abs/1512.06493}{{\ttfamily
  arXiv:1512.06493 [hep-ph]}}.

\bibitem{Ballon-Bayona:2017sxa}
A.~Ballon-Bayona, H.~Boschi-Filho, L.~A.~H. Mamani, A.~S. Miranda, and V.~T.
  Zanchin, ``{Effective holographic models for QCD: glueball spectrum and trace
  anomaly},'' \href{http://dx.doi.org/10.1103/PhysRevD.97.046001}{{\em Phys.
  Rev. D} {\bfseries 97} no.~4, (2018) 046001},
  \href{http://arxiv.org/abs/1708.08968}{{\ttfamily arXiv:1708.08968
  [hep-th]}}.

\bibitem{Ballon-Bayona:2017bwk}
A.~Ballon-Bayona, G.~Krein, and C.~Miller, ``{Strong couplings and form factors
  of charmed mesons in holographic QCD},''
  \href{http://dx.doi.org/10.1103/PhysRevD.96.014017}{{\em Phys. Rev. D}
  {\bfseries 96} no.~1, (2017) 014017},
  \href{http://arxiv.org/abs/1702.08417}{{\ttfamily arXiv:1702.08417
  [hep-ph]}}.

\bibitem{Ballon-Bayona:2020qpq}
A.~Ballon-Bayona and L.~A.~H. Mamani, ``{Nonlinear realization of chiral
  symmetry breaking in holographic soft wall models},''
  \href{http://dx.doi.org/10.1103/PhysRevD.102.026013}{{\em Phys. Rev. D}
  {\bfseries 102} no.~2, (2020) 026013},
  \href{http://arxiv.org/abs/2002.00075}{{\ttfamily arXiv:2002.00075
  [hep-ph]}}.

\bibitem{Ballon-Bayona:2021ibm}
A.~Ballon-Bayona, L.~A.~H. Mamani, and D.~M. Rodrigues, ``{Spontaneous chiral
  symmetry breaking in holographic soft wall models},''
  \href{http://dx.doi.org/10.1103/PhysRevD.104.126029}{{\em Phys. Rev. D}
  {\bfseries 104} no.~12, (2021) 126029},
  \href{http://arxiv.org/abs/2107.10983}{{\ttfamily arXiv:2107.10983
  [hep-ph]}}.

\bibitem{Gubser:2008yx}
S.~S. Gubser, A.~Nellore, S.~S. Pufu, and F.~D. Rocha, ``{Thermodynamics and
  bulk viscosity of approximate black hole duals to finite temperature quantum
  chromodynamics},''
  \href{http://dx.doi.org/10.1103/PhysRevLett.101.131601}{{\em Phys. Rev.
  Lett.} {\bfseries 101} (2008) 131601},
  \href{http://arxiv.org/abs/0804.1950}{{\ttfamily arXiv:0804.1950 [hep-th]}}.

\bibitem{Gursoy:2008za}
U.~Gursoy, E.~Kiritsis, L.~Mazzanti, and F.~Nitti, ``{Holography and
  Thermodynamics of 5D Dilaton-gravity},''
  \href{http://dx.doi.org/10.1088/1126-6708/2009/05/033}{{\em JHEP} {\bfseries
  05} (2009) 033}, \href{http://arxiv.org/abs/0812.0792}{{\ttfamily
  arXiv:0812.0792 [hep-th]}}.

\bibitem{Bigazzi:2009bk}
F.~Bigazzi, A.~L. Cotrone, J.~Mas, A.~Paredes, A.~V. Ramallo, and J.~Tarrio,
  ``{D3-D7 Quark-Gluon Plasmas},''
  \href{http://dx.doi.org/10.1088/1126-6708/2009/11/117}{{\em JHEP} {\bfseries
  11} (2009) 117}, \href{http://arxiv.org/abs/0909.2865}{{\ttfamily
  arXiv:0909.2865 [hep-th]}}.

\bibitem{Bigazzi:2011it}
F.~Bigazzi, A.~L. Cotrone, J.~Mas, D.~Mayerson, and J.~Tarrio, ``{D3-D7
  Quark-Gluon Plasmas at Finite Baryon Density},''
  \href{http://dx.doi.org/10.1007/JHEP04(2011)060}{{\em JHEP} {\bfseries 04}
  (2011) 060}, \href{http://arxiv.org/abs/1101.3560}{{\ttfamily arXiv:1101.3560
  [hep-th]}}.

\bibitem{He:2013qq}
S.~He, S.-Y. Wu, Y.~Yang, and P.-H. Yuan, ``{Phase Structure in a Dynamical
  Soft-Wall Holographic QCD Model},''
  \href{http://dx.doi.org/10.1007/JHEP04(2013)093}{{\em JHEP} {\bfseries 04}
  (2013) 093}, \href{http://arxiv.org/abs/1301.0385}{{\ttfamily arXiv:1301.0385
  [hep-th]}}.

\bibitem{Ballon-Bayona:2013cta}
A.~Ballon-Bayona, ``{Holographic deconfinement transition in the presence of a
  magnetic field},'' \href{http://dx.doi.org/10.1007/JHEP11(2013)168}{{\em
  JHEP} {\bfseries 11} (2013) 168},
  \href{http://arxiv.org/abs/1307.6498}{{\ttfamily arXiv:1307.6498 [hep-th]}}.

\bibitem{Alho:2013hsa}
T.~Alho, M.~J\"arvinen, K.~Kajantie, E.~Kiritsis, C.~Rosen, and K.~Tuominen,
  ``{A holographic model for QCD in the Veneziano limit at finite temperature
  and density},'' \href{http://dx.doi.org/10.1007/JHEP04(2014)124}{{\em JHEP}
  {\bfseries 04} (2014) 124}, \href{http://arxiv.org/abs/1312.5199}{{\ttfamily
  arXiv:1312.5199 [hep-ph]}}. [Erratum: JHEP 02, 033 (2015)].

\bibitem{Bigazzi:2014qsa}
F.~Bigazzi and A.~L. Cotrone, ``{Holographic QCD with Dynamical Flavors},''
  \href{http://dx.doi.org/10.1007/JHEP01(2015)104}{{\em JHEP} {\bfseries 01}
  (2015) 104}, \href{http://arxiv.org/abs/1410.2443}{{\ttfamily arXiv:1410.2443
  [hep-th]}}.

\bibitem{Dudal:2014jfa}
D.~Dudal and T.~G. Mertens, ``{Melting of charmonium in a magnetic field from
  an effective AdS/QCD model},''
  \href{http://dx.doi.org/10.1103/PhysRevD.91.086002}{{\em Phys. Rev. D}
  {\bfseries 91} (2015) 086002},
  \href{http://arxiv.org/abs/1410.3297}{{\ttfamily arXiv:1410.3297 [hep-th]}}.

\bibitem{Rougemont:2015oea}
R.~Rougemont, R.~Critelli, and J.~Noronha, ``{Holographic calculation of the
  QCD crossover temperature in a magnetic field},''
  \href{http://dx.doi.org/10.1103/PhysRevD.93.045013}{{\em Phys. Rev. D}
  {\bfseries 93} no.~4, (2016) 045013},
  \href{http://arxiv.org/abs/1505.07894}{{\ttfamily arXiv:1505.07894
  [hep-th]}}.

\bibitem{Knaute:2017opk}
J.~Knaute, R.~Yaresko, and B.~K\"ampfer, ``{Holographic QCD phase diagram with
  critical point from Einstein\textendash{}Maxwell-dilaton dynamics},''
  \href{http://dx.doi.org/10.1016/j.physletb.2018.01.053}{{\em Phys. Lett. B}
  {\bfseries 778} (2018) 419--425},
  \href{http://arxiv.org/abs/1702.06731}{{\ttfamily arXiv:1702.06731
  [hep-ph]}}.

\bibitem{Critelli:2017oub}
R.~Critelli, J.~Noronha, J.~Noronha-Hostler, I.~Portillo, C.~Ratti, and
  R.~Rougemont, ``{Critical point in the phase diagram of primordial
  quark-gluon matter from black hole physics},''
  \href{http://dx.doi.org/10.1103/PhysRevD.96.096026}{{\em Phys. Rev. D}
  {\bfseries 96} no.~9, (2017) 096026},
  \href{http://arxiv.org/abs/1706.00455}{{\ttfamily arXiv:1706.00455
  [nucl-th]}}.

\bibitem{Ballon-Bayona:2017dvv}
A.~Ballon-Bayona, M.~Ihl, J.~P. Shock, and D.~Zoakos, ``{A universal order
  parameter for Inverse Magnetic Catalysis},''
  \href{http://dx.doi.org/10.1007/JHEP10(2017)038}{{\em JHEP} {\bfseries 10}
  (2017) 038}, \href{http://arxiv.org/abs/1706.05977}{{\ttfamily
  arXiv:1706.05977 [hep-th]}}.

\bibitem{Gursoy:2017wzz}
U.~Gursoy, M.~Jarvinen, and G.~Nijs, ``{Holographic QCD in the Veneziano Limit
  at a Finite Magnetic Field and Chemical Potential},''
  \href{http://dx.doi.org/10.1103/PhysRevLett.120.242002}{{\em Phys. Rev.
  Lett.} {\bfseries 120} no.~24, (2018) 242002},
  \href{http://arxiv.org/abs/1707.00872}{{\ttfamily arXiv:1707.00872
  [hep-th]}}.

\bibitem{Braga:2018zlu}
N.~R.~F. Braga and L.~F. Ferreira, ``{Heavy meson dissociation in a plasma with
  magnetic fields},''
  \href{http://dx.doi.org/10.1016/j.physletb.2018.06.053}{{\em Phys. Lett. B}
  {\bfseries 783} (2018) 186--192},
  \href{http://arxiv.org/abs/1802.02084}{{\ttfamily arXiv:1802.02084
  [hep-ph]}}.

\bibitem{Arefeva:2020byn}
I.~Y. Aref'eva, K.~Rannu, and P.~Slepov, ``{Holographic anisotropic model for
  light quarks with confinement-deconfinement phase transition},''
  \href{http://dx.doi.org/10.1007/JHEP06(2021)090}{{\em JHEP} {\bfseries 06}
  (2021) 090}, \href{http://arxiv.org/abs/2009.05562}{{\ttfamily
  arXiv:2009.05562 [hep-th]}}.

\bibitem{Rannu:2022fxw}
K.~Rannu, I.~Y. Aref\textquoteright{}eva, and P.~S. Slepov, ``{Holographic
  model in anisotropic hot dense QGP with external Magnetic Field},''
  \href{http://dx.doi.org/10.31349/SuplRevMexFis.3.0308126}{{\em Rev. Mex. Fis.
  Suppl.} {\bfseries 3} no.~3, (2022) 0308126}.

\bibitem{Arefeva:2022avn}
I.~Y. Aref'eva, A.~Ermakov, K.~Rannu, and P.~Slepov, ``{Holographic model for
  light quarks in anisotropic hot dense QGP with external magnetic field},''
  \href{http://dx.doi.org/10.1140/epjc/s10052-022-11166-3}{{\em Eur. Phys. J.
  C} {\bfseries 83} no.~1, (2023) 79},
  \href{http://arxiv.org/abs/2203.12539}{{\ttfamily arXiv:2203.12539
  [hep-th]}}.

\bibitem{Grefa:2022fpu}
J.~Grefa, M.~Hippert, J.~Noronha, J.~Noronha-Hostler, I.~Portillo, C.~Ratti,
  and R.~Rougemont, ``{QCD Equilibrium and Dynamical Properties from
  Holographic Black Holes},''
  \href{http://dx.doi.org/10.31349/SuplRevMexFis.3.040910}{{\em Rev. Mex. Fis.
  Suppl.} {\bfseries 3} no.~4, (2022) 040910},
  \href{http://arxiv.org/abs/2207.12564}{{\ttfamily arXiv:2207.12564
  [nucl-th]}}.

\bibitem{Grefa:2021qvt}
J.~Grefa, J.~Noronha, J.~Noronha-Hostler, I.~Portillo, C.~Ratti, and
  R.~Rougemont, ``{Hot and dense quark-gluon plasma thermodynamics from
  holographic black holes},''
  \href{http://dx.doi.org/10.1103/PhysRevD.104.034002}{{\em Phys. Rev. D}
  {\bfseries 104} no.~3, (2021) 034002},
  \href{http://arxiv.org/abs/2102.12042}{{\ttfamily arXiv:2102.12042
  [nucl-th]}}.

\bibitem{dePaula:2020bte}
W.~de~Paula, C.-R. Ji, J.~P. B.~C. de~Melo, T.~Frederico, and O.~Louren\c{c}o,
  ``{The holographic paradigm of hadron dynamics for medium modified nuclear
  matters},'' \href{http://dx.doi.org/10.1016/j.physletb.2020.135339}{{\em
  Phys. Lett. B} {\bfseries 803} (2020) 135339}.

\bibitem{Li:2023mpv}
Z.~Li, J.~Liang, S.~He, and L.~Li, ``{Holographic study of higher-order baryon
  number susceptibilities at finite temperature and density},''
  \href{http://arxiv.org/abs/2305.13874}{{\ttfamily arXiv:2305.13874
  [hep-ph]}}.

\bibitem{Li:2014dsa}
D.~Li, S.~He, and M.~Huang, ``{Temperature dependent transport coefficients in
  a dynamical holographic QCD model},''
  \href{http://dx.doi.org/10.1007/JHEP06(2015)046}{{\em JHEP} {\bfseries 06}
  (2015) 046}, \href{http://arxiv.org/abs/1411.5332}{{\ttfamily arXiv:1411.5332
  [hep-ph]}}.

\bibitem{Li:2014hja}
D.~Li, J.~Liao, and M.~Huang, ``{Enhancement of jet quenching around phase
  transition: result from the dynamical holographic model},''
  \href{http://dx.doi.org/10.1103/PhysRevD.89.126006}{{\em Phys. Rev. D}
  {\bfseries 89} no.~12, (2014) 126006},
  \href{http://arxiv.org/abs/1401.2035}{{\ttfamily arXiv:1401.2035 [hep-ph]}}.

\bibitem{Ballon-Bayona:2018ddm}
A.~Ballon-Bayona, H.~Boschi-Filho, L.~A.~H. Mamani, A.~S. Miranda, and V.~T.
  Zanchin, ``{An effective holographic approach to QCD},'' in {\em {14th
  International Workshop on Hadron Physics}}.
\newblock 4, 2018.
\newblock \href{http://arxiv.org/abs/1804.01579}{{\ttfamily arXiv:1804.01579
  [hep-th]}}.

\bibitem{Ballon-Bayona:2021tzw}
A.~Ballon-Bayona, L.~A.~H. Mamani, A.~S. Miranda, and V.~T. Zanchin,
  ``{Effective holographic models for QCD: Thermodynamics and viscosity
  coefficients},'' \href{http://dx.doi.org/10.1103/PhysRevD.104.046013}{{\em
  Phys. Rev. D} {\bfseries 104} no.~4, (2021) 046013},
  \href{http://arxiv.org/abs/2103.14188}{{\ttfamily arXiv:2103.14188
  [hep-th]}}.

\bibitem{Mamani:2019mgu}
L.~A.~H. Mamani, ``{Conformal symmetry breaking in holographic QCD},''
  \href{http://dx.doi.org/10.1103/PhysRevD.100.106009}{{\em Phys. Rev. D}
  {\bfseries 100} no.~10, (2019) 106009},
  \href{http://arxiv.org/abs/1910.00026}{{\ttfamily arXiv:1910.00026
  [hep-th]}}.

\bibitem{Cai:2012xh}
R.-G. Cai, S.~He, and D.~Li, ``{A hQCD model and its phase diagram in
  Einstein-Maxwell-Dilaton system},''
  \href{http://dx.doi.org/10.1007/JHEP03(2012)033}{{\em JHEP} {\bfseries 03}
  (2012) 033}, \href{http://arxiv.org/abs/1201.0820}{{\ttfamily arXiv:1201.0820
  [hep-th]}}.

\bibitem{Cremonini:2012ny}
S.~Cremonini, U.~Gursoy, and P.~Szepietowski, ``{On the Temperature Dependence
  of the Shear Viscosity and Holography},''
  \href{http://dx.doi.org/10.1007/JHEP08(2012)167}{{\em JHEP} {\bfseries 08}
  (2012) 167}, \href{http://arxiv.org/abs/1206.3581}{{\ttfamily arXiv:1206.3581
  [hep-th]}}.

\bibitem{Mamani:2020pks}
L.~A.~H. Mamani, C.~V. Flores, and V.~T. Zanchin, ``{Phase diagram and compact
  stars in a holographic QCD model},''
  \href{http://dx.doi.org/10.1103/PhysRevD.102.066006}{{\em Phys. Rev. D}
  {\bfseries 102} no.~6, (2020) 066006},
  \href{http://arxiv.org/abs/2006.09401}{{\ttfamily arXiv:2006.09401
  [hep-th]}}.

\bibitem{Ballon-Bayona:2020xls}
A.~Ballon-Bayona, H.~Boschi-Filho, E.~F. Capossoli, and D.~M. Rodrigues,
  ``{Criticality from Einstein-Maxwell-dilaton holography at finite temperature
  and density},'' \href{http://dx.doi.org/10.1103/PhysRevD.102.126003}{{\em
  Phys. Rev. D} {\bfseries 102} no.~12, (2020) 126003},
  \href{http://arxiv.org/abs/2006.08810}{{\ttfamily arXiv:2006.08810
  [hep-th]}}.

\bibitem{Gursoy:2007cb}
U.~Gursoy and E.~Kiritsis, ``{Exploring improved holographic theories for QCD:
  Part I},'' \href{http://dx.doi.org/10.1088/1126-6708/2008/02/032}{{\em JHEP}
  {\bfseries 02} (2008) 032}, \href{http://arxiv.org/abs/0707.1324}{{\ttfamily
  arXiv:0707.1324 [hep-th]}}.

\bibitem{Kinar:1998vq}
Y.~Kinar, E.~Schreiber, and J.~Sonnenschein, ``{Q anti-Q potential from strings
  in curved space-time: Classical results},''
  \href{http://dx.doi.org/10.1016/S0550-3213(99)00652-5}{{\em Nucl. Phys. B}
  {\bfseries 566} (2000) 103--125},
  \href{http://arxiv.org/abs/hep-th/9811192}{{\ttfamily arXiv:hep-th/9811192}}.

\bibitem{Iatrakis:2010jb}
I.~Iatrakis, E.~Kiritsis, and A.~Paredes, ``{An AdS/QCD model from tachyon
  condensation: II},'' \href{http://dx.doi.org/10.1007/JHEP11(2010)123}{{\em
  JHEP} {\bfseries 11} (2010) 123},
  \href{http://arxiv.org/abs/1010.1364}{{\ttfamily arXiv:1010.1364 [hep-ph]}}.

\bibitem{Workman:2022ynf}
{\bfseries Particle Data Group} Collaboration, R.~L. Workman {\em et~al.},
  ``{Review of Particle Physics},''
  \href{http://dx.doi.org/10.1093/ptep/ptac097}{{\em PTEP} {\bfseries 2022}
  (2022) 083C01}.

\bibitem{OBELIX:1997zla}
{\bfseries OBELIX} Collaboration, A.~Bertin {\em et~al.}, ``{Study of anti-p p
  --\ensuremath{>} 2pi+ 2pi- annihilation from S states},''
  \href{http://dx.doi.org/10.1016/S0370-2693(97)01189-1}{{\em Phys. Lett. B}
  {\bfseries 414} (1997) 220--228}.

\bibitem{Zyla:2020zbs}
{\bfseries Particle Data Group} Collaboration, P.~Zyla {\em et~al.}, ``{Review
  of Particle Physics},'' \href{http://dx.doi.org/10.1093/ptep/ptaa104}{{\em
  PTEP} {\bfseries 2020} no.~8, (2020) 083C01}.

\bibitem{Donoghue:1992dd}
J.~F. Donoghue, E.~Golowich, and B.~R. Holstein,
  \href{http://dx.doi.org/10.1017/CBO9780511524370}{{\em {Dynamics of the
  standard model}}}, vol.~2.
\newblock CUP, 2014.

\bibitem{Gokalp:2001}
O.~Y. A.~Gokalp, Y.~Sarac, ``{Scalar $a_0$-meson contributions to radiative
  $\omega\rightarrow \pi^{0}\eta\gamma$ and $\rho^{0}\rightarrow
  \pi^{0}\eta\gamma$decays},''
  \href{http://dx.doi.org/10.1007/s100520100800}{{\em Eur. Phys. J. C}
  {\bfseries 22} (2001) 327}.

\bibitem{Isgur:1988vm}
N.~Isgur, C.~Morningstar, and C.~Reader, ``{The a1 in tau Decay},''
  \href{http://dx.doi.org/10.1103/PhysRevD.39.1357}{{\em Phys. Rev. D}
  {\bfseries 39} (1989) 1357}.

\bibitem{Shuryak:1988ck}
E.~V. Shuryak, \href{http://dx.doi.org/10.1142/0161}{{\em {The QCD vacuum,
  hadrons and the superdense matter}}}, vol.~8.
\newblock World Scientific Lecture Notes in Physics: Volume 71, 1988.

\bibitem{Giusti:1998wy}
L.~Giusti, F.~Rapuano, M.~Talevi, and A.~Vladikas, ``{The QCD chiral condensate
  from the lattice},''
  \href{http://dx.doi.org/10.1016/S0550-3213(98)00659-2}{{\em Nucl. Phys. B}
  {\bfseries 538} (1999) 249--277},
  \href{http://arxiv.org/abs/hep-lat/9807014}{{\ttfamily
  arXiv:hep-lat/9807014}}.

\bibitem{Fukaya:2009fh}
{\bfseries JLQCD} Collaboration, H.~Fukaya, S.~Aoki, S.~Hashimoto, T.~Kaneko,
  J.~Noaki, T.~Onogi, and N.~Yamada, ``{Determination of the chiral condensate
  from 2+1-flavor lattice QCD},''
  \href{http://dx.doi.org/10.1103/PhysRevLett.104.122002}{{\em Phys. Rev.
  Lett.} {\bfseries 104} (2010) 122002},
  \href{http://arxiv.org/abs/0911.5555}{{\ttfamily arXiv:0911.5555 [hep-lat]}}.
  [Erratum: Phys.Rev.Lett. 105, 159901 (2010)].

\bibitem{McNeile:2012xh}
C.~McNeile, A.~Bazavov, C.~T.~H. Davies, R.~J. Dowdall, K.~Hornbostel, G.~P.
  Lepage, and H.~D. Trottier, ``{Direct determination of the strange and light
  quark condensates from full lattice QCD},''
  \href{http://dx.doi.org/10.1103/PhysRevD.87.034503}{{\em Phys. Rev. D}
  {\bfseries 87} no.~3, (2013) 034503},
  \href{http://arxiv.org/abs/1211.6577}{{\ttfamily arXiv:1211.6577 [hep-lat]}}.

\bibitem{Miranda:2009uw}
A.~S. Miranda, C.~A. Ballon~Bayona, H.~Boschi-Filho, and N.~R.~F. Braga,
  ``{Black-hole quasinormal modes and scalar glueballs in a finite-temperature
  AdS/QCD model},'' \href{http://dx.doi.org/10.1088/1126-6708/2009/11/119}{{\em
  JHEP} {\bfseries 2009} no.~11, (2009) 119},
  \href{http://arxiv.org/abs/0909.1790}{{\ttfamily arXiv:0909.1790 [hep-th]}}.

\bibitem{Mamani:2013ssa}
L.~A.~H. Mamani, A.~S. Miranda, H.~Boschi-Filho, and N.~R.~F. Braga, ``{Vector
  meson quasinormal modes in a finite-temperature AdS/QCD model},''
  \href{http://dx.doi.org/10.1007/JHEP03(2014)058}{{\em JHEP} {\bfseries 2014}
  no.~3, (2014) 1--26}, \href{http://arxiv.org/abs/1312.3815}{{\ttfamily
  arXiv:1312.3815 [hep-th]}}.

\bibitem{Bartz:2016ufc}
S.~P. Bartz and T.~Jacobson, ``{Chiral Phase Transition and Meson Melting from
  AdS/QCD},'' \href{http://dx.doi.org/10.1103/PhysRevD.94.075022}{{\em Phys.
  Rev. D} {\bfseries 94} (2016) 075022},
  \href{http://arxiv.org/abs/1607.05751}{{\ttfamily arXiv:1607.05751
  [hep-ph]}}.

\bibitem{Mamani:2018uxf}
L.~A.~H. Mamani, A.~S. Miranda, and V.~T. Zanchin, ``{Melting of scalar mesons
  and black-hole quasinormal modes in a holographic QCD model},''
  \href{http://dx.doi.org/10.1140/epjc/s10052-019-6902-5}{{\em Eur. Phys. J. C}
  {\bfseries 79} no.~5, (2019) 1--20},
  \href{http://arxiv.org/abs/1809.03508}{{\ttfamily arXiv:1809.03508
  [hep-th]}}.

\bibitem{Mamani:2022qnf}
L.~A.~H. Mamani, D.~Hou, and N.~R.~F. Braga, ``{Melting of heavy vector mesons
  and quasinormal modes in a finite density plasma from holography},''
  \href{http://dx.doi.org/10.1103/PhysRevD.105.126020}{{\em Phys. Rev. D}
  {\bfseries 105} no.~12, (2022) 126020},
  \href{http://arxiv.org/abs/2204.08068}{{\ttfamily arXiv:2204.08068
  [hep-ph]}}.

\end{thebibliography}\endgroup

\end{document}